\documentclass[a4paper,11pt]{article}    
\pdfoutput=1
\usepackage[OT1]{fontenc} 
\usepackage{multicol}
\usepackage{graphicx}  
\usepackage{bbold} 
\usepackage{mathtools}
\usepackage{amsmath}
\numberwithin{equation}{section} 
\usepackage{amsfonts} 
\usepackage{amssymb}
\usepackage{amsbsy}
\usepackage{amsfonts}
\usepackage{physics}
\usepackage{bbold}
\usepackage{slashed}  
\usepackage{color}
\usepackage[table]{xcolor}
\usepackage{tablestyles}
\usepackage{tabularx}
\usepackage{hyperref}
\usepackage{orcidlink}
\usepackage{bbding}

\definecolor{oucrimsonred}{rgb}{0.6, 0.0, 0.0}
\definecolor{DarkGray}{gray}{0.4}
\definecolor{forestgreen}{rgb}{0.13,0.35,0.13}
\definecolor{ocre}{HTML}{F16723}

\usepackage[page,toc,titletoc,title]{appendix}

\numberwithin{equation}{section}
\numberwithin{table}{section}
\numberwithin{figure}{section}
\usepackage{bm}
\usepackage{cite}  
\usepackage{mathrsfs}
\usepackage{framed}
\usepackage[font=footnotesize,labelfont=bf]{caption}
\def\eq#1{{Eq.~(\ref{#1})}}
\def\eqs#1#2{{Eqs.~(\ref{#1})--(\ref{#2})}}

\def\Tr{\mbox{Tr}\,}

\colorlet{grayline}{gray!70}
\definecolor{blueline}{rgb}{0,0.27,0.55}
\definecolor{DarkGray}{gray}{0.4}
\definecolor{Gray}{gray}{0.6}
\definecolor{oucrimsonred}{rgb}{0.6, 0.0, 0.0}
\definecolor{persianblue}{rgb}{0.11, 0.22, 0.73}
\definecolor{forestgreen}{rgb}{0.13,0.35,0.13}
 \hypersetup{colorlinks, citecolor=forestgreen, linkcolor=forestgreen, urlcolor=forestgreen}
\newcommand{\be}{\begin{equation}}
\newcommand{\ee}{\end{equation}}
\newcommand{\bea}{\begin{eqnarray}}
\newcommand{\eea}{\end{eqnarray}}
\newcommand{\nn}{\nonumber}

\newcommand{\CC}{\operatorname{C}}
\newcommand{\BB}{\operatorname{B}}

\newcommand*\xbar[1]{%
  \hbox{\;%
    \vbox{%
      \hrule height 0.5pt 
      \kern0.5ex
      \hbox{%
        \kern-0.25em
        \ensuremath{#1}%
        \kern-0.07em
      }%
    }%
  }%
} 
\newcommand{\com}[1]{}
\newcommand{\gsim}{\lower.7ex\hbox{$\;\stackrel{\textstyle>}{\sim}\;$}}
\newcommand{\lsim}{\lower.7ex\hbox{$\;\stackrel{\textstyle<}{\sim}\;$}} 

\newcommand{\bc}{\begin{center}}
\newcommand{\ec}{\end{center}}
\newcommand{\cmb}{\mathscr{C}_2}

\font\beeg=cmr17 scaled 1800
\newbox\ibox
\def\versal#1{\setbox\ibox=\hbox{{\beeg #1}~}%
	    \noindent\global\hangindent=\wd\ibox\global\hangafter-2%
	    \sc\smash{\llap {\lower 14pt \box\ibox}}}
\begin{document}

\hypersetup{citecolor = forestgreen,
linktoc = section, 
linkcolor = forestgreen, 
urlcolor = forestgreen
}

\thispagestyle{empty}
\begin{center}
{ \Large \color{oucrimsonred} \textbf{ 
 Bell inequality is violated 
 in charmonium decays}}

\vspace*{1.5cm}
{\color{DarkGray}
  {\bf M. Fabbrichesi$^{a \orcidlink{0000-0003-1937-3854}}$,}
{\bf   R. Floreanini$^{a \orcidlink{0000-0002-0424-2707}}$,}
{\bf E. Gabrielli$^{b,a,c, \orcidlink{0000-0002-0637-5124}}$ and} 
 {\bf L. Marzola$^{c,d \orcidlink{0000-0003-2045-1100}}$}
}\\

\vspace{0.5cm}
{\small 
{\it  \color{DarkGray} (a)
INFN, Sezione di Trieste, Via Valerio 2, \\I-34127 Trieste, Italy}
\\[1mm]
  {\it \color{DarkGray}
    (b) Physics Department, University of Trieste, Strada Costiera 11, \\ I-34151 Trieste, Italy}
  \\[1mm]  
  {\it \color{DarkGray}
(c) Laboratory of High-Energy and Computational Physics, NICPB, R\"avala 10, \\ 10143 Tallinn, Estonia}
}  \\[1mm]
  {\it \color{DarkGray}
(d) Institute of Computer Science, University of Tartu, Narva mnt 18, \\ 51009 Tartu, Estonia.
}
\ec

 \vskip0.5cm
\bc
{\color{DarkGray}
\rule{0.7\textwidth}{0.5pt}}
\ec
\vskip1cm
\bc
{\bf ABSTRACT} 
\ec
\noindent The experimental data on the helicity amplitudes  of  charmonium decays  allow us to measure entanglement in final state spin correlations and test possible violations of the Bell inequality.  We find that the Bell inequality is violated with a significance of  5$\sigma$ or more  in  the decays 
\begin{eqnarray}
 \eta_{c}, \, \chi_{c}^{0}, \, J/\psi  &\to&  \Lambda +\bar \Lambda \nonumber \\
J/\psi & \to & \Xi^- +\bar \Xi^+,\; \Xi^0 +\bar \Xi^0, \;\Sigma^{-} +\bar \Sigma^{+},\;  \Sigma^{0} +\bar \Sigma^{0}, \nonumber \\
\psi(3686)& \to & \Xi^- +\bar \Xi^+,\;   \Sigma^- +\bar \Sigma^+,\;  \Sigma^{0} +\bar \Sigma^{0}, \nonumber \\
\chi^{0}_{c}, \, \chi_{c}^{1} & \to & \phi+ \phi \, .\nonumber  
\end{eqnarray}
The decays $\psi(3686) \to  \Lambda +\bar \Lambda$ and $\Xi^0 +\bar \Xi^0$ show the same violation but with less significance. The decay  $\psi(3686) \to  \Omega^- + \bar \Omega^+$ displays entanglement. These results firmly establish the presence of entanglement and  quantum non-separability  at high energies, in a setting with  particles of different spins  and interacting through electroweak and strong interactions. In addition, the relatively long lifetime of some of the strange baryons produced in the decays provides a natural probe to test whether quantum spin correlations remain  after the particles have interacted with the beam pipe and the first few layers of the detector. 

\vspace*{5mm}

\noindent

  \vskip 3cm
\bc 
{\color{DarkGray} 
\SquareShadowBottomRight
}
\ec

	\newpage

	\tableofcontents
	
	

\newpage
\section{Introduction\label{sec:intro}} 

{\versal  The violation of the Bell inequality~\cite{Bell:1987hh}}  shows that quantum mechanics cannot be explained in terms of  local  variables with  definite properties that are independent of their measurement.
It has been verified experimentally at low energies~\cite{Aspect:1982fx,Weihs:1998gy} by means of
two photons prepared into a singlet state, whose  polarizations  are measured along different directions to verify their entanglement~\cite{Horodecki:2009zz} and the  violation of  the Bell inequality.

Entanglement and Bell inequality violation have been  first established in high-energy physics in $B$-meson decays~\cite{Fabbrichesi:2023idl}. Entanglement, but not yet Bell inequality violation, has also been observed in top-quark production~\cite{ATLAS:2023fsd, CMS:2024pts}. These quantum  observables are now actively investigated in high-energy physics and the reader can find most of the pertinent references in the recent review paper~\cite{Barr:2024djo}.

The charmonium decays have been singled out early on in~\cite{Tornqvist:1980af,Tornqvist:1986pe} and \cite{Baranov:2008zzb,Baranov:2009zza}  as  promising systems in which to check the violation of Bell inequality  in particle physics. The  charmonium decays $\eta_{c} \to \Lambda\bar  \Lambda$,   $\chi_{c} \to \Lambda\bar  \Lambda$ and $J/\psi \to  \Lambda \bar \Lambda$ were originally discussed and further studied in \cite{Chen:2013epa}. 

In recent years, data on the helicity amplitudes for  these processes have been published by the BESIII Collaboration~\cite{BESIII:2018cnd,BESIII:2022qax}, which has also studied the  similar $J/\psi \to \Sigma^- \bar \Sigma^+$\cite{BESIII:2020fqg}, $\Sigma^{0}\bar \Sigma^{0}$~\cite{BESIII:2024nif}, $ \Xi^-\bar\Xi^+$~\cite{BESIII:2021ypr} or $\Xi^0 \bar \Xi^0$~\cite{BESIII:2023drj} and $\psi(3686) \to \Sigma^- \bar \Sigma^+$\cite{BESIII:2020fqg},  $\Sigma^{0}\bar \Sigma^{0}$~\cite{BESIII:2024nif}, $ \Xi^-\bar\Xi^+$~\cite{BESIII:2022lsz} or $\Xi^0 \bar \Xi^0$~\cite{BESIII:2023lkg} decays, as well as $\psi(3686) \to  \Omega^-\bar\Omega^+$~\cite{BESIII:2020lkm}. In addition, the three decays $ \chi_{c}^{J} \to \phi \phi$ (with $J=0,\, 1,\, 2$) have also been analyzed~\cite{BESIII:2023zcs}. The ATLAS~\cite{ATLAS:2014swk}, CMS~\cite{CMS:2018wjk} and LHC$b$~\cite{LHCb:2013hzx} collaborations also provided data on the decay $\Lambda_{b}\to J/\psi \, \Lambda$, which has a charmonium state in the final state.  
Entanglement is often mentioned in describing the final state of charmonium decays but only as a descriptive feature and no attempt of a quantitative study  has been attempted yet.

The helicity amplitudes provided by the experimental collaborations with dedicated data analyses allow to perform a full quantum tomography of the system of interest, reconstructing the density matrix that describes the polarizations and spin correlations of the final state. The amount of entanglement present in the latter and the violation of Bell inequality can then be straightforwardly investigated by means of various operators, as described in Section 2. The uncertainty of the measurements  determines, through the propagation of the errors, the  significance of the presence of entanglement and the violation of the Bell inequality. We implement this program with the limitation that, for some of the decays, the phases of the helicity amplitudes are not given and that the correlations among the uncertainties are not always available. Notice that the phases in the amplitude can only originate from the strong interactions among the final states  because the absorptive part from the  weak interactions is either loop suppressed or vanishing. 

The same experimental data on the helicity amplitudes can also be used to test to what extent the density matrix factorizes into momentum and spin dependent parts: The entanglement of the strange baryons is computed after some of them  have crossed the beam pipe outer wall and the first few layers of the detector  and, therefore, a comparison with the expected theoretical value can tell us whether entanglement has been affected by these interactions or not. On the other hand, we know that the momentum dependent part of the density matrix has completely lost coherence since we can see the tracks left by the particles~\cite{mott1929}. This is an important test that probes, for the first time, the full density matrix of the system as a whole.

\subsection{The charmonium system}

\begin{figure}[ht!]
\begin{center}
\includegraphics[width=5in]{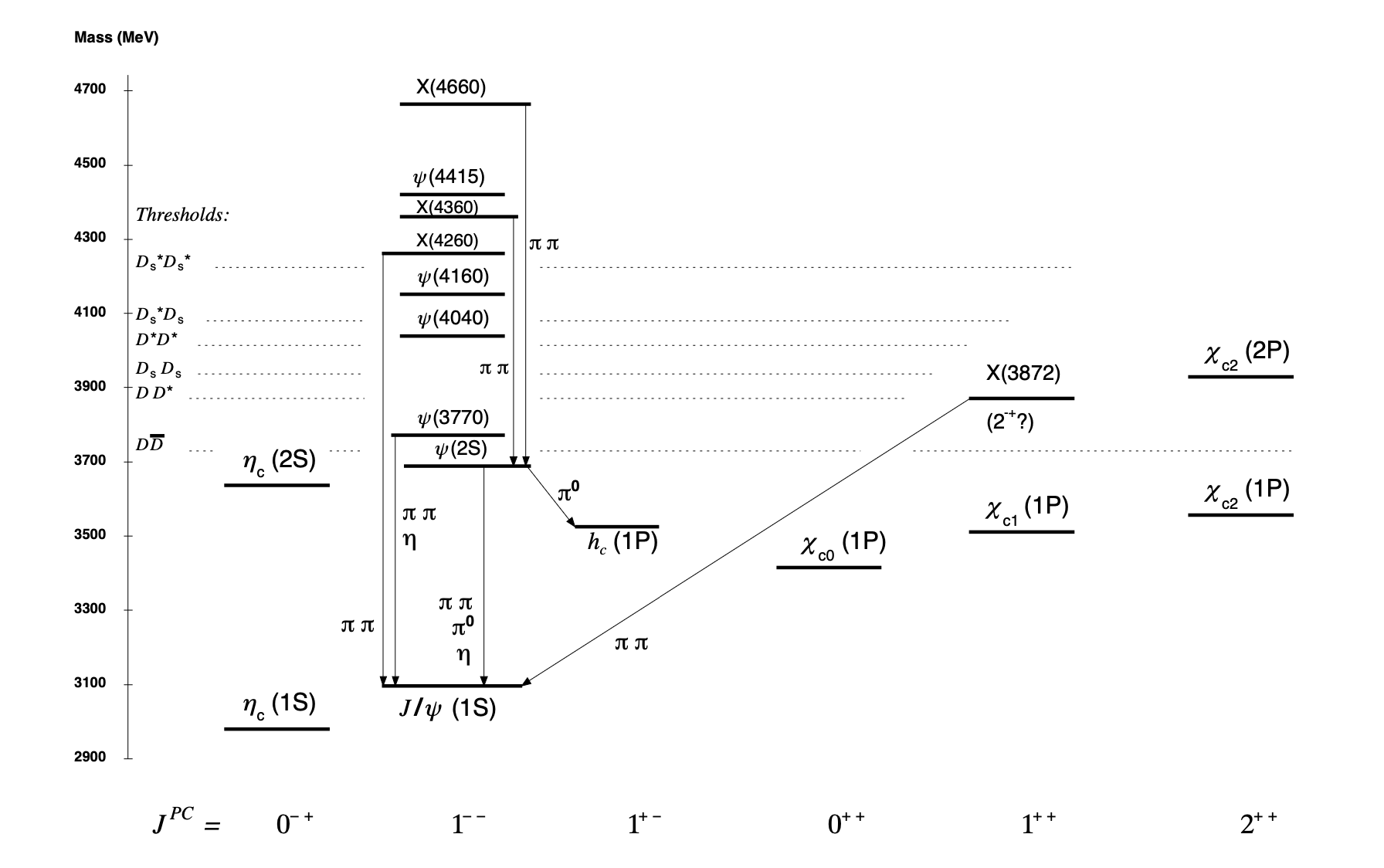}. \caption{\footnotesize The charmonium system. The bottom line shows the spin and parities of the states. Figure  from~\cite{ParticleDataGroup:2022pth} [\href{https://creativecommons.org/licenses/by/4.0/}{CC BY 4.0}]
\label{fig:charmonium} 
}
\end{center}
\end{figure}

{\versal  The charmonium system} encompasses a rich variety of bound states with a structure resembling that of an atomic system. Figure~\ref{fig:charmonium} shows some of these states and lists their properties. They present themselves at colliders as resonances whose decays can be studied in detail.
 
These states and their decays are a natural laboratory for the study  of the entanglement in spin correlations at high energies. Charmonium gives access to a variety of bipartite final states characterized by different spins and to initial resonant states that encompass, as well, spins ranging from 0 to 2.
 
The Beijing electron-positron collider (BEPCII) and spectrometer (BESIII) have been designed to operate in the charmonium energy regime (between 2 and 4.95 GeV) with a peak  luminosity of $10^{33}$ cm$^{-2}$s$^{-1}$. Electron and positron are collided at these energies, charmonium states are produced in isolation or in combination with a photon, and their decay products recorded in the detector. The BESIII detector~\cite{BESIII:2009fln} consists of a multilayered drift chamber (MDC), a plastic scintillator time-of-flight system (TOF) and an electromagnetic calorimeter (EMC). The entire detector is enclosed in a superconducting solenoid magnet providing a 1 T magnetic field.
  
Many charmonium states have been copiously produced in the last few years at BEPCII, thus making it possible not only to measure their branching fractions into different final states, but also the helicity amplitudes of each single decay. These are reconstructed by means of the angular distribution of the decay products: protons, pions and kaons. The full quantum tomography of the bipartite final states uses all the information provided by the charmonium two-body decays and encodes it into the polarization density matrix.

\section{Methods \label{sec:methods}} 

{\versal Since our goal is to utilize} the experimental values of helicity amplitudes to directly find entanglement and test the violation of the Bell inequality, we do not need to compute the polarization density matrix from a Lagrangian of a specific model. The representations of the elements of the $SO(3)$ group are sufficient to parameterize the density matrix in terms of the helicity amplitudes
$w_{\lambda_{1}\lambda_{2}}$ provided by the experiments~\cite{Jacob:1959at,Berestetskii:1982qgu,Leader:2011vwq}:
 \be
\rho_{\lambda_{1}\lambda_{2},\lambda_{1}^{\prime}\lambda_{2}^{\prime}}\propto w_{\lambda_{1}\lambda_{2}}w^{*}_{\lambda_{1}^{\prime}\lambda_{2}^{\prime}} \sum_{k} D^{(J)*}_{k,\lambda_{1}-\lambda_{2}}(0,\Theta,0) D^{(J)}_{k,\lambda_{1}^{\prime}-\lambda_{2}^{\prime}}(0,\Theta,0) \, ,\label{rho}
\ee
where $D^{(J)}_{i,j}$ is the Wigner $D$-matrix for the spin $J$ of the decaying state and $k$ runs over all the possible helicity values of the same state.  
The overall factor in \eq{rho} is set by the normalization requirement that $\Tr \rho=1$. The density matrix in \eq{rho} is written in the center-of-mass reference frame where the momenta of final state particles are equal in magnitude and opposite in direction and, therefore, the total helicity of the two-particle system is $\lambda_1-\lambda_2$. The dependence on the angle $\phi$ drops out in the products of the Wigner matrices because of the cylindrical symmetry of the problem:
\be
D^{(J)*}_{k,\lambda_{1}-\lambda_{2}}(0,\Theta,0) D^{(J)}_{k,\lambda_{1}^{\prime}-\lambda_{2}^{\prime}}(0,\Theta,0)=D^{(J)*}_{k,\lambda_{1}-\lambda_{2}}(\phi,\Theta,0) D^{(J)}_{k,\lambda_{1}^{\prime}-\lambda_{2}^{\prime}}(\phi,\Theta,0) \, .
\ee

Depending on the symmetries enjoyed by a decay process, the number of independent helicity amplitudes required for a full description may be reduced: it is zero for decays of scalar states into fermions, 2 for the same decays into spin 1 states and for the decay of a vector state into two fermions, four for the case of the decay into two spin 3/2 fermions.

In general, the number of independent amplitudes is reduced by imposing helicity conservation, that is
\be
|\lambda_1 - \lambda_2| \leq J
\ee
for the decay $A\to 1+2$, with $J$ the spin of the particle $A$.

A further reduction in the number of independent helicity amplitudes comes from parity conservation, which implies
\be
w^J_{\lambda_1,\lambda_2} = \eta_A\, \eta_1\,\eta_2 (-1)^{J-s_1-s_2} \, w^J_{-\lambda_1,-\lambda_2} \, ,
\ee
in which $\eta_i$ are the intrinsic parities and $s_i$ the spin of the particles in the final state.

For final states including identical particles, helicity amplitudes transform  under the interchange of the particles as  
\be
w^J_{\lambda_1,\lambda_2} = (-1)^{J-2 s} \, w^J_{\lambda_2,\lambda_1}  \, ,
\ee
with $s=s_1=s_2$.  If instead the final state is made of a pair of particle and anti-particle:
\be
w^J_{\lambda_1,\lambda_2} = \eta_C \, (-1)^{J} \, w^J_{\lambda_2,\lambda_1}  \, ,
\ee
in which  $\eta_C$ is the $C$ parity of the decaying particle $A$.


\subsection{Tools to study entanglement and  test the violation of Bell inequality}

The determination of the density matrix is the aim of quantum tomography. In the present case we find the polarization density matrix from the analysis of the experimental data as presented by the experimental collaborations.

The density matrix makes it possible to compute the entanglement and test Bell inequalities for the final states of the charmonium decays. The choice of the most appropriate tools depends on whether the final state is described by qubits (two-level systems)
or qutrits (three-level systems) or, more in general, qudits ($d$-level systems).

\subsubsection{Qubits}

Consider a bipartite system composed by a spin-1/2 pair, one controlled by an observer, Alice, and the other by a second observer, Bob. The corresponding quantum state can be described by a $4 \times 4$
density matrix of the form:
\begin{equation}
\rho=\frac{1}{4}\Big[\mathbb{1}_2\otimes\mathbb{1}_2 + \sum_{i=1}^3 \BB^{+}_i (\sigma_i\otimes\mathbb{1}_2)
+ \sum_{i=1}^3 \BB^-_j (\mathbb{1}_2\otimes \sigma_j) + \sum_{i,j=1}^3 \CC_{ij} (\sigma_i\otimes\sigma_j) \Big]\ ,
\label{rho-1/2}
\end{equation}
where $\sigma_i$ are the Pauli matrices, $\mathbb{1}_2$ is the unit $2\times 2$ matrix and the indices $i$ and $j$ running over $1$, $2$, $3$, represent any three orthogonal spatial directions.

The real coefficients $\BB^+_i $ and $\BB_j^{-}$
represent the polarization of the two spin-1/2 fermions, while the real matrix $\CC_{ij} $ gives their spin correlations. The density matrix in (\ref{rho-1/2}) is normalized, ${\rm Tr}[\rho]=1$,
while extra constraints on $\BB^+_i$, $\BB^-_i$
and $\CC_{ij}$
need to be enforced to guarantee its positivity, as all eigenvalues of a density matrix are necessarily non-negative.

The entanglement content of any bipartite system described with the density matrix $\rho$, that is, a measure of
the amount of quantum correlations among the two composing sub-systems, can be quantified with the concurrence
$\mathscr{C}[\rho]$, taking values between zero (for separable, unentangled states)
and 1 (maximally entangled states).
In the case of two spin-1/2 sub-systems, a two qubit system, the concurrence can be analytically computed through the auxiliary matrix
\begin{equation} 
R=\rho \,  (\sigma_y \otimes \sigma_y) \, \rho^* \, (\sigma_y \otimes \sigma_y)\, ,
\label{auxiliary-R}
\end{equation}
where $\rho^*$ denotes a matrix with complex conjugated entries. Although non-Hermitian, the matrix $R$
possesses non-negative eigenvalues; denoting $r_i$, $i=1,2,3,4$, their square roots
and assuming  $r_1$ to be  the largest,
the concurrence of the state $\rho$ can be expressed as~\cite{Wootters:PhysRevLett.80.2245}
\begin{equation}
\mathscr{C}[\rho] = \max \big( 0, r_1-r_2-r_3-r_4 \big)\ .
\label{concurrence}
\end{equation}

In quantum mechanics a statistical language is adopted for the description of physical phenomena. Interestingly, this compelling tool
is amenable to experimental verification against alternative, fully deterministic, local description of natural phenomena through Bell locality tests.

In the case of a two spin-1/2 system, Alice and Bob are assumed to measure
two spin-observable each, ($\hat A_1$, $\hat A_2$), and ($\hat B_1$, $\hat B_2$),
typically spin projections along four different unit vectors,
$\vec{n}_1$, $\vec{n}_3$ for Alice, and $\vec{n}_2$, $\vec{n}_4$ for Bob, so that
$\hat{A}_1= \vec{n}_1\cdot \vec{\sigma}$ and similarly for the remaining three observables. The Bell test consists in determining the following combination of joint expectation values
\begin{equation}
\mathcal{I}_2=\langle {\hat A}_1 {\hat B}_1\rangle + \langle {\hat A}_1 {\hat B}_2\rangle +\langle {\hat A}_2 {\hat B}_1\rangle
- \langle {\hat A}_2 {\hat B}_2\rangle\ ,
\label{CHSH}
\end{equation}
that in any, local, deterministic model cannot exceed a value of 2.
In quantum mechanics, $\mathcal{I}_2$ can be conveniently expressed as an expectation of a Bell operator $\mathscr{B}_2$, $\mathcal{I}_2={\rm Tr}[\rho\,  \mathscr{B}_2]$, where
\begin{equation}\label{Bchsh}
\mathscr{B}_2 ={\vec n}_1 \cdot {\vec\sigma} \otimes ({\vec n}_2 - {\vec n}_4) \cdot {\vec\sigma}
+ {\vec n}_3 \cdot {\vec\sigma} \otimes ({\vec n}_2 + {\vec n}_4 )\cdot  {\vec\sigma}\ .
\end{equation}
If an actual experiment finds $\mathcal{I}_2 > 2$, one has to deduce that some sort of nonlocal
resource had been shared between the two parties, and this is precisely what is predicted by quantum mechanics.

In practice, given an experimentally collected correlation data, one thus needs to maximize $\mathcal{I}_2$ in~(\ref{CHSH}) by choosing
suitable four independent spatial directions.
Fortunately, this optimization process can be performed in full generality
for a generic spin correlation matrix~\cite{Horodecki:1995340}.
Indeed, consider the matrix $\CC$ and its transpose $\CC^T$
and form the symmetric, positive, $3\times 3$ matrix
$M= \CC \CC^T$; its three eigenvalues $m_1$, $m_2$, $m_3$ can be ordered in increasing order:
$m_1\geq m_2\geq m_3$. Then, the two-spin state $\rho$ in (\ref{rho-1/2}) violates the Bell inequality $\mathcal{I}_2\leq 2$
if and only if the sum of the two greatest eigenvalues of $M$ is strictly larger than 1, that is (Horodecki condition)
\begin{equation}
\mathfrak{m}_{12}\equiv m_1 + m_2 >1\, .
\label{eigenvalue-inequality}
\end{equation}
%

\subsubsection{Qutrits}

The density operator representing the state of a bipartite system made of two qutrits is a $9\times 9$ matrix
that can be written as
\be
\rho = \frac{1}{9}\left[\mathbb{1}_3\otimes
   \mathbb{1}_3\right]+
   \sum_{a=1}^8 f_a \left[T^a\otimes \mathbb{1}_3\right]+\sum_{a=1}^8 g_a \left[\mathbb{1}_3\otimes T^a\right]
   +\sum_{a,b=1}^8 h_{ab}  \left[T^a\otimes T^b\right]\, ,
\label{eq:rho-qutrit}
\ee
where $T^a$ are the standard Gell-Mann matrices, while $\mathbb{1}_3$ is the unit $3\times 3$ matrix.

Although an analytic expression for the concurrence of a generic two-qutrit state is lacking, a lower bound on its value can be given in terms of the single spin polarizations coefficients,
$f_a$ and $g_a$, and the correlation matrix $h_{ab}$ appearing
in the decomposition (\ref{eq:rho-qutrit}):
\bea
\cmb &=& 2\max \Big[ -\frac{2}{9}-12 \sum_a f_{a}^{2} +6 \sum_a g_{a}^{2} + 4 \sum_{ab} h_{ab}^{2}\, ;\nn \Big.\\
 & & \hskip 2cm \Big. -\frac{2}{9}-12  \sum_a g_{a}^{2} +6 \sum_a f_{a}^{2} + 4 \sum_{ab} h_{ab}^{2},\; 0 \Big]\,.
\label{eq:cmb-qutrit}
\eea

As in the case of qubits, a Bell test for a system of two qutrits results in the determination
of a combination ${\cal I}_{3}$ of joint expectations values involving
four spin observables, ($\hat A_1$, $\hat A_2$) for Alice,
and ($\hat B_1$, $\hat B_2$) for Bob.
In quantum mechanics, it can be again expressed as an expectation value on the state (\ref{eq:cmb-qutrit})
of a suitable Bell operator $\mathscr{B}_3$:
\begin{equation}
{\cal I}_{3}={\rm Tr}\big[\rho\, \mathscr{B}_3\big] \ .
\label{eq:I3}
\end{equation}
The explicit form of $\mathscr{B}_3$ depends on the choice of the four measured operators $\hat{A}_1$, $\hat{A}_2$,
$\hat{B}_1$, $\hat{B}_2$.
For the case of the maximally correlated qutrit state,
the problem of finding an optimal choice of measurements has been solved~\cite{Collins:PhysRevLett.88.040404}, and the Bell operator takes a particular simple form~\cite{Latorre:PhysRevA.65.052325}:
\be
\small
\mathscr{B}_3 =  \begin{pmatrix}
 0 & 0 & 0 & 0 & 0 & 0 & 0 & 0 & 0  \\
 0 & 0 & 0 & -\dfrac{2}{\sqrt{3}} & 0 & 0& 0 & 0 & 0  \\
 0 & 0 &0 & 0 & -\dfrac{2}{\sqrt{3}} & 0 &2 & 0 & 0  \\
 0 &  -\dfrac{2}{\sqrt{3}} & 0 & 0 & 0 & 0 & 0 &0 & 0  \\
 0& 0 & -\dfrac{2}{\sqrt{3}} & 0 & 0 & 0 & -\dfrac{2}{\sqrt{3}} & 0 &0  \\
 0 & 0 & 0 & 0 & 0 & 0 & 0 &  -\dfrac{2}{\sqrt{3}} & 0  \\
 0 & 0 & 2 & 0 & -\dfrac{2}{\sqrt{3}} & 0 &0& 0 & 0  \\
 0 & 0 & 0 & 0 & 0 &  -\dfrac{2}{\sqrt{3}} & 0 & 0 & 0  \\
 0 & 0 & 0 & 0 & 0& 0 & 0 & 0 & 0  \\
\end{pmatrix} \ .
\label{B}
\ee

Within the choice of measurements leading to the Bell operator (\ref{B}),
there is still the freedom of modifying the measured observables through local unitary transformations,
which effectively corresponds to local changes of basis, separately at Alice's and Bob's sites.
Correspondingly, the Bell operator undergoes the change:
\begin{equation}
\mathscr{B}_3 \to (U\otimes V)^{\dag} \cdot \mathscr{B}_3\cdot (U\otimes V)\ , \label{uni_rot}
\end{equation}
where $U$ and $V$ are independent $3\times 3$ unitary matrices.
One can use this additional freedom in order to maximize the value of ${\cal I}_3$ for any given bipartite qutrit state described by $\rho$.

\subsubsection{Additional observables}

Given a density matrix $\rho$ describing the state of a generic bipartite system, $S_A + S_B$,
the reduced density matrix describing the state of $S_A$ alone is given by
$\rho_A={\rm Tr}_B[\rho]$, where the trace is performed on all $S_B$ degrees of freedom;
similarly,  $\rho_B={\rm Tr}_A[\rho]$ is the reduced density matrix describing the state of $S_B$.

For pure states, $\rho = |\psi\rangle\langle \psi|$, or equivalently $\rho^2 = \rho$, the quantity
\begin{equation}
\mathscr{E}[\rho] \equiv -{\rm Tr}[\rho_A\ln \rho_A] = -{\rm Tr}[\rho_B\ln \rho_B]\ ,
\label{entropy}
\end{equation}
giving the von Neumann entropy of the reduced density matrices, is a good entanglement quantifier---often called in the literature entropy of entanglement. Indeed, a pure state $\rho$ is entangled if and only
if its reduced density matrices have non-zero entropy.
Assuming the two systems $S_A$ and $S_B$ have the same dimension $d$,
one finds $0\leq \mathscr{E}[\rho] \leq \ln d$;
the first equality holds if and only if the bipartite pure state is separable,
while the upper bound is reached by a maximally entangled state.

Given a generic density matrix $\rho$ for the bipartite state $S_A + S_B$,
deciding whether the state is entangled or not, or quantifying its entanglement content is in general a hard problem~\cite{Gurvits:2003gdo,Gharibian:2008hgo},
and, thus, one has to  rely on quantities that give only sufficient conditions for the presence of entanglement.

One such quantity involves the operation of partial transposition.
Given a basis of orthonormal vectors $\{ |ij\rangle=|i\rangle\otimes| j\rangle\}$ for the system $S_A + S_B$,
any density matrix can be represented by its matrix elements
$\langle i_1j_1|\rho|i_2 j_2 \rangle$; then, the partially transpose matrix $\rho^{T_B}$ with respect to $S_B$
is represented by matrix elements $\langle i_1j_2|\rho|i_2 j_1 \rangle$; a similar expression holds for $\rho^{T_A}$.
Interestingly, if $\rho^{T_B}$, or equivalently $\rho^{T_A}$ possesses negative eigenvalues, than the composite system is entangled. In addition, the absolute sum of the negative eigenvalues
of $\rho^{T_B}$, called negativity,
\be
{\cal N}(\rho) = \sum_k \frac{|\lambda_k| - \lambda_k}{2}\, ,
\ee
$\lambda_k$ being the eigenvalues of $\rho^{T_B}$,
can be used to quantify its entanglement content~\cite{Vidal:2002zz}.

\subsection{From the  helicity amplitudes to the quantum observables}

The helicity amplitudes are extracted from the data by means of a maximum likelihood fit. This fit  depends on all the kinematic variables of the processes under consideration. For instance, for the case of the decay of $J/\psi$ into  $\Lambda$-$\bar \Lambda$ baryon pairs, the differential cross section $\cal W$ dependence can be indicated as 
\be
{\cal W} (\alpha, \Delta\Phi, \alpha_\Lambda, \alpha_{\bar \Lambda},\Theta_\Lambda, \boldsymbol{n}_1, \boldsymbol{n}_2 ) \label{xsec}
\ee
in which $\alpha$ and $\Delta\Phi$ are directly related to the parametrization of the helicity amplitudes---and are therefore not varied, $\alpha_\Lambda$, $\alpha_{\bar \Lambda}$ are the polarimetric coefficients of the baryons, $ \boldsymbol{n}_i$ are the directions of the proton ($i=1$)  and antiproton ($i=2$) in the rest frame of the $J/\psi$ and $\Theta_\Lambda$ is the scattering angle. The likelihood fit yields the coefficients $\alpha$ and $\Delta\Phi$ and therefore the helicity amplitudes, which are quantities independent of the other kinematic variables in \eq{xsec}. 

In this work, we use the values of the helicity amplitudes thus obtained by inserting them in the quantum observables that quantify entanglement and the violation of Bell inequality. The uncertainties associated to the helicity amplitudes can be propagated to find the uncertainties of the utilized operators.

\section{Charmonium spin 0 states}

{\versal The decays of the spin 0 states} of the charmonium are the simplest to analyze because the density matrix only depends on one or, at most, two helicity amplitudes. Moreover, the density matrix is independent of the scattering angle.

\subsection[$\eta_c \to \Lambda + \bar  \Lambda\quad \text{and} \quad \chi^0_c \to \Lambda + \bar  \Lambda$]{$\boxed{\;\; \eta_c \to \Lambda + \bar  \Lambda\quad \text{and} \quad \chi^0_c \to \Lambda + \bar  \Lambda\;\; }$ }

The scalar and pseuodoscalar   states of the charmonium can decay into a pair of strange $\Lambda$ baryon and anti-baryon
\be
\eta_c   \to \Lambda +  \bar \Lambda \quad \text{and} \quad \chi_c^0  \to \Lambda +  \bar \Lambda \,,
\ee
with branching fraction $(1.10\pm0.28)\times 10^{-3}$ and  $(1.27\pm0.09)\times 10^{-4}$~\cite{ParticleDataGroup:2022pth}, respectively. The scalar state $\eta_c$ is produced in the processes
\be
e^+\, e^- \to J/\psi \to \gamma\, \eta_c\, ,
\ee
while the  $\chi^0_c$ in 
\be
e^{+} \, ^{-} \to \psi (3686) \to \gamma \,\chi^{0} \, .
\ee

The $\Lambda$ baryons and anti-baryons decay into $p\, \pi^-$ and $\bar p\, \pi^-$, respectively. The angular dependence of these charged final states in the cascade decays allows the reconstruction of the baryon polarizations.

The final states are constrained---by the conservation of the helicity---to be described by the state
\be
|\psi_0\rangle  \propto  w_{ \frac{1}{2}\,- \frac{1}{2}} \, |\tfrac{1}{2}, \tfrac{1}{2} \rangle \otimes |\tfrac{1}{2}, -\tfrac{1}{2} \rangle +
w_{-\frac{1}{2}\,\frac{1}{2} }\, |\tfrac{1}{2}, -\tfrac{1}{2} \rangle \otimes |\tfrac{1}{2}, \tfrac{1}{2} \rangle \label{state0}
\ee
in which $w_{ij}$ are the helicity amplitudes and $|J,\, m\rangle$ the spin states.
Parity sets the relative sign between the two amplitudes: it is $-1$ for the pseudoscalar $\eta_{c}$ and $1$ for the scalar  $\chi_c^{0} $. Accordingly, the $\eta_c $ falls into the singlet representation of the product $\tfrac{1}{2} \otimes \tfrac{1}{2} = 0 \oplus 1$ while the $\chi_c^0$ into the $m=0$ component of the triplet. Charge parity conservation implies the same condition as parity and does not add new relations among the helicity amplitudes.

The  states in \eq{state0}  give the following density matrix
\be
\rho_{\Lambda\,\Lambda}  =  |\psi_0\rangle \langle \psi_0| = \frac{1}{2}\,  \begin{pmatrix} 
0&0&0&0\\
0& 1 & \pm1 &0\\
0&\pm 1&1 &0\\
0&0&0&0 
\end{pmatrix} \, ,
\label{rho0}
\ee
in which the only, still undefined,  overall  size of the amplitudes has canceled out in the normalization, which is $\Tr \rho =1$.
 
The system is completely constrained, thus implementing the idealized two-qubit system of textbooks. This property was already observed for the  decay of the Higgs boson $H\to \tau^{-}\tau^{+}$ in \cite{Fabbrichesi:2022ovb}. Neither the $\Lambda$ baryon nor the anti-baryon  are  polarized.

The concurrence can be computed and it is maximal:
\be
\mathscr{C} [\rho]=1\, .
\ee

From the density matrix in \eq{rho0}, using the Pauli matrices, we can write  the correlation matrix
\be 
C_{ij} =\Tr\left(\rho_{\Lambda \Lambda} \,\sigma_{i} \otimes \sigma_{j}\right) =\begin{pmatrix} 1 &0 &0\\ 0&1&0\\ 0&0&-1 \end{pmatrix} \, ,
\ee
which is the same for both the decay processes. Accordingly, the Horodecki condition is found to be
\be
  \mathfrak{m}_{12} =2 \, ,
\ee
corresponding to a maximal violation of the Bell inequality.

For these decays, we do not even need the experimental values of the helicity amplitudes to claim maximum entanglement and Bell inequality violation. Uncertainties from the data analysis are however necessary to assess the significance of the result. These values have not yet been released by the experimental collaboration. In any case, if the uncertainty turns out to be of the same order of that of the processes discussed below, that is, of the per mille, the violation of the Bell inequality will easily be established with a significance of more than 100$\sigma$.

This process provides a direct test of the conservation of  quantum correlations. If the experiments find  a difference between the helicity amplitudes  $w_{ \frac{1}{2}\,- \frac{1}{2}}$ and  $w_{ -\frac{1}{2}\, \frac{1}{2}}$, or that they vanish,  it will mean that some of the original coherence has been lost during the  flight of the $\Lambda$ baryons---some of which  travel inside the beam pipe wall and the first layers of the detector before decaying. This is an important test, as explained in the Introduction. We come back to this point in Section 6.

\subsection[$\chi^{0}_{c}\to\phi+ \phi$]{$\boxed{\;\; \chi^{0}_{c}\to\phi+ \phi\;\;}$}

The scalar   state of the charmonium can decay into a pair of $\phi$ mesons
\be
\chi^{0}_c   \to \phi + \phi \,,
\ee
with branching fraction of $(8.48\pm0.26 \pm 0.27)\times 10^{-4}$~\cite{BESIII:2023zcs}.

The $\chi^{0}_c$ are produced, as already mentioned, in
\be
e^{+} e^{-} \to \psi (3686) \to \gamma \chi^{0} \, .
\ee

The final state of the two $\phi$ mesons can  be written as 
\be
|\Psi \rangle =w_{_{-1\, -1} }\, |-1,\, -1\rangle + w_{_{0\,0}}\, |0\, 0 \rangle+  w_{_{1\,1}}\, |1,\, 1\rangle  \, ,\label{pure}
\ee
with
\be
 |w_{_{-1\, -1}}|^2 + |w_{_{0\,0}}|^2 + |w_{_{1\,1}} |^2 =1 \, ,
\ee
and $w_{_{1 \, 1}} = -w_{_{-1\, -1}}$ because of the conservation of parity. The same condition is found by the indistinguishability of the final state particles. There is therefore only one independent amplitude and the density matrix depends on one complex number.

The final states are spin 1 and their polarizations are described by qutrits. 
The resulting $9\times 9$ density matrix $\rho_{\phi\phi} = |\Psi \rangle \langle \Psi |$ is written as
\be
\small
\rho_{\phi\phi} \propto  \begin{pmatrix} 
  0 & 0 & 0 & 0 & 0 & 0 & 0 & 0 & 0  \\
  0 & 0 & 0 & 0 & 0 & 0 & 0 & 0 & 0  \\
  0 & 0 &  |w_{_{-1\, -1}}|^2 & 0 & w_{_{-1\, -1}} w_{_{0\, 0}}^*& 0 &  w_{_{-1\, -1}} w_{_{1\, 1}}^*& 0 & 0  \\
  0 & 0 & 0 & 0 & 0 & 0 & 0 & 0 & 0  \\
  0 & 0 & w_{_{0\, 0}} w_{_{-1, -1}}^*& 0 &|w_{_{0\,0}}|^2  & 0 & w_{_{0\, 0}} w_{_{1\, 1}}^*& 0 & 0  \\
  0 & 0 & 0 & 0 & 0 & 0 & 0 & 0 & 0  \\
  0 & 0 &  w_{_{1\, 1} } w_{_{-1\, -1}}^*& 0 &  w_{_{1\, 1}} w_{_{0\, 0}}^*& 0 &   |w_{_{1\, 1}}|^2& 0 & 0  \\
  0 & 0 & 0 & 0 & 0 & 0 & 0 & 0 & 0  \\
  0 & 0 & 0 & 0 & 0 & 0 & 0 & 0 & 0  \\
\end{pmatrix} \, ,
\label{rhochiphiphi}
\ee
The analysis of the data in~\cite{BESIII:2023zcs}  selects $2701\pm84$ out of the $\gamma K^+K^-K^+K^-$ final states events. The maximum likelihood fit yields the absolute value of the ratio of the moduli of the helicity amplitudes:
\be
\left| \frac{w_{_{1, 1}}}{w_{_{0\, 0}}}\right|=0.299\pm0.003|_{\rm stat} \, \pm 0.019|_{\rm syst} \, . \label{exp1}
\ee
No value for the relative phase is provided. Accordingly, we can only carry out the analysis in the case of zero phase. As pointed out in the Introduction, this phase comes from the final state strong interactions if we assume that the form factors have no significant absorptive part.

The entanglement can be determined from the entropy of entanglement given in \eq{entropy}  because the final state in \eq{pure} is pure. We find, after propagating the errors,
\be
 \mathscr{E}[\rho] = 0.531\pm 0.040\,.
\ee
This number differs from zero with a significance of 13.3$\sigma$.

After optimization, the expectation value of the Bell operator  is
\be
\Tr  \rho_{\phi\phi} \, \mathscr{B}  = 2.296 \pm 0.034\, .
\ee

This decay provides a clean test of the violation of Bell inequality in a system of two qutrits. Its significance is $8.8\sigma$.

\section{Charmonium spin 1 states}\label{sec:1to1212}
{\versal The decay of spin 1 particles} brings in a dependence of the polarization density matrix on the scattering angle. The amount of entanglement and possible violations of the Bell inequality therefore depend on the value of this angle. 

Data on many different processes are available and we review all of them. Such a comprehensive  presentation is  necessarily repetitive. We apologize. The final results are summarized in Table~\ref{tab:fermions}.

\subsection[$J/\psi \to \Lambda + \bar  \Lambda \quad \text{and} \quad \psi(3686) \to  \Lambda + \bar  \Lambda $]{$\boxed{J/\psi \to \Lambda + \bar  \Lambda \quad \text{and} \quad \psi(3686) \to  \Lambda + \bar  \Lambda }$}

The helicity states of the final system in
\be
J/\psi  \to \Lambda +  \bar \Lambda
\ee
fall in the  triplet representation of the product $\tfrac{1}{2} \otimes \tfrac{1}{2} = 1 \oplus 0$. It is constrained by the conservation of the angular momentum to be described by the three states
\bea
|\psi_{\uparrow}\rangle &\propto & w_{\frac{1}{2}\, \frac{1}{2}} \,| \tfrac{1}{2}\, \tfrac{1}{2} \rangle \otimes | \tfrac{1}{2}\, \tfrac{1}{2} \rangle\nn\\
|\psi_{\downarrow}\rangle &\propto & w_{-\frac{1}{2}\, -\frac{1}{2}} \,| \tfrac{1}{2}\, -\tfrac{1}{2} \rangle \otimes | \tfrac{1}{2}\, -\tfrac{1}{2} \rangle\nn\\
|\psi_{0}\rangle &\propto & w_{\frac{1}{2}\, -\frac{1}{2}} \,| \tfrac{1}{2}\, \tfrac{1}{2} \rangle \otimes | \tfrac{1}{2}\, -\tfrac{1}{2} \rangle + w_{-\frac{1}{2}\, \frac{1}{2}} \,| \tfrac{1}{2}\, -\tfrac{1}{2} \rangle \otimes | \tfrac{1}{2}\, \tfrac{1}{2} \rangle  \, , \label{states3}
\eea
in which the state in the first line of \eq{states3} corresponds to the $J/\psi$ being transversally polarized with positive helicity ($J_z=+1$), the second line to the opposite helicity ($J_z=-1)$ and the third line to the 0 helicity ($J_{z}=0$), that is, the $J/\psi$ being longitudinally polarized. The states in \eq{states3} are written along the $z$-axis and must be rotated to the direction of the final state momenta.

In the process
\be
e^{+}e^{-} \to\gamma \to  c \bar c \to J/\psi \to \Lambda \bar \Lambda \, ,
\ee
the $J/\psi$ is produced polarized. The correlation matrix of the two baryons depends on the scattering angle $\Theta$ because the polarization of the $J/\psi$ does.

The elements of the density matrix can be written as 
\be
\rho_{\lambda_{1}\lambda_{2},\lambda_{1}^{\prime}\lambda_{2}^{\prime}}\propto w_{\lambda_{1}\lambda_{2}}w^{*}_{\lambda_{1}^{\prime}\lambda_{2}^{\prime}} \sum_{k=\pm1} D^{(1)*}_{k,\lambda_{1}-\lambda_{2}}(0,\Theta,0) D^{(1)}_{k,\lambda_{1}^{\prime}-\lambda_{2}^{\prime}}(0,\Theta,0)
\ee
where $D^{(1)}_{i,j}$ is the Wigner matrix for the spin 1 representation of $SO(3)$ and the sum is only over the  $\pm1$ polarizations because the spin 1 state is produced from unpolarized electrons and positrons with the electron and positron taken to be massless and, therefore, with only the  $\pm1$ helicities.

Of the four helicity amplitudes, only two are independent.  The density matrix is given by
\be
\small
\rho_{\Lambda\bar\Lambda} \propto 
\begin{pmatrix}  w_{_{-\frac{1}{2}\, -\frac{1}{2}}}  w_{_{-\frac{1}{2}\,- \frac{1}{2}}}^{*} s_\Theta^2 &-w_{_{-\frac{1}{2}\, -\frac{1}{2}}}  w_{_{-\frac{1}{2}\,\frac{1}{2}}}^{*} \frac{c_\Theta s_\Theta}{\sqrt{2}}& w_{_{-\frac{1}{2}\, -\frac{1}{2}}}  w_{_{\frac{1}{2}\,- \frac{1}{2}}}^{*} \frac{c_\Theta s_\Theta}{\sqrt{2}} &w_{_{-\frac{1}{2}\, -\frac{1}{2}}}  w_{_{\frac{1}{2}\,\frac{1}{2}}}^{*}  s_\Theta^2  \\
-w_{_{-\frac{1}{2}\, \frac{1}{2}}}  w_{_{-\frac{1}{2}\,- \frac{1}{2}}}^{*} \frac{c_\Theta s_\Theta}{\sqrt{2}}&w_{_{-\frac{1}{2}\, \frac{1}{2}}}  w_{_{-\frac{1}{2}\,\frac{1}{2}}}^{*} f_\Theta& w_{_{-\frac{1}{2}\, \frac{1}{2}}}  w_{_{\frac{1}{2}\,- \frac{1}{2}}}^{*}\frac{ s_\Theta^2}{2} &-w_{_{-\frac{1}{2}\, \frac{1}{2}}}  w_{_{\frac{1}{2}\,\frac{1}{2}}}^{*} \frac{c_\Theta s_\Theta}{\sqrt{2}}\\
w_{_{\frac{1}{2}\,- \frac{1}{2}}}  w_{_{-\frac{1}{2}\,- \frac{1}{2}}}^{*}\frac{c_\Theta s_\Theta}{\sqrt{2}}&w_{_{\frac{1}{2}\, -\frac{1}{2}}}  w_{_{-\frac{1}{2}\,\frac{1}{2}}}^{*}  \frac{ s_\Theta^2}{2} & w_{_{\frac{1}{2}\, -\frac{1}{2}}}  w_{_{\frac{1}{2}\,- \frac{1}{2}}}^{*} f_\Theta&w_{_{\frac{1}{2}\, -\frac{1}{2}}}  w_{_{\frac{1}{2}\,\frac{1}{2}}}^{*} \frac{c_\Theta s_\Theta}{\sqrt{2}}\\
w_{_{\frac{1}{2}\, \frac{1}{2}}}  w_{_{-\frac{1}{2}\,- \frac{1}{2}}}^{*} s_\Theta^2  &-w_{_{\frac{1}{2}\, \frac{1}{2}}}  w_{_{-\frac{1}{2}\,\frac{1}{2}}}^{*} \frac{c_\Theta s_\Theta}{\sqrt{2}}& w_{_{\frac{1}{2}\, \frac{1}{2}}}  w_{_{\frac{1}{2}\,- \frac{1}{2}}}^{*}\frac{c_\Theta s_\Theta}{\sqrt{2}}&w_{_{\frac{1}{2}\, \frac{1}{2}}}  w_{_{\frac{1}{2}\,\frac{1}{2}}}^{*} s_\Theta^2   \label{rhoLambda}
\end{pmatrix}\, ,
\ee
in which $f_\Theta \equiv (3 - \cos 2 \Theta )/4$, $s_\Theta \equiv \sin \Theta$ and $c_\Theta \equiv \cos \Theta$.

The helicity amplitudes can be  parametrized as
 \be
 w_{\frac{1}{2}\, \frac{1}{2}}=w_{-\frac{1}{2}\, -\frac{1}{2}}=\frac{\sqrt{1-\alpha}}{\sqrt{2}}\quad \text{ and} \quad 
  w_{\frac{1}{2}\, -\frac{1}{2}} = w_{-\frac{1}{2}\, \frac{1}{2}}=\sqrt{1+\alpha} \, \exp [-i \Delta \Phi ]\,.
  \ee
  
  The polarization of the $\Lambda$ baryons is given by
\be
B_i^-=-B^+_i= \Tr \rho_{\Lambda \Lambda}  \mathbb{1} \otimes \sigma_{i} =(0,\frac{ \sqrt{1- \alpha^{2}} \sin 2  \Theta  \sin \Delta \Phi}{C_{0}},0)\, , \label{pol}
 \ee
in which $C_0=2 + \alpha + \alpha  \cos 2 \Theta$. The expression for the  polarization in \eq{pol}   agrees with \cite{Perotti:2018wxm,BESIII:2022qax}.
  
 Ten billion $J/\psi$ events have been collected at the BESIII detector. The decay $J/\psi \to \Lambda \bar \Lambda$ has branching fraction $(1.89 \pm 0.08) \times 10^{-3}$~\cite{ParticleDataGroup:2022pth}. The decay into $\Lambda\bar \Lambda$ pairs is reconstructed from their dominant hadron decays: $\Lambda \to p\pi^-$ and $\bar \Lambda \to \bar p \pi^-$. The maximum likelihood fit yields the values of the two parameters defining the helicity amplitudes~\cite{BESIII:2022qax}:
  \be
 \alpha= 0.4748 \pm 0.0022 |_{\rm stat}\pm 0.0031|_{\rm syst} \quad \text{and} \quad \Delta \Phi= 0.7521\pm0.0042|_{\rm stat} \pm 0.0066|_{\rm syst}\, .
 \ee 
No correlation in the uncertainties is provided.

The spin correlation matrix can be computed from the density matrix in \eq{rhoLambda}  and
it  is given by
\be
\CC= \frac{1}{C_{0}}\,\begin{pmatrix}   2\,\sin^{2}\Theta  & 0 & \sqrt{1-\alpha^{2}}\sin 2 \, \Theta \cos \Delta\Phi \\
0 &2\, \alpha  \sin^{2}\Theta   &0 \\
- \sqrt{1- \alpha^{2}} \sin 2 \,\Theta \cos \Delta\Phi & 0 & -(1+ 2 \alpha +\cos 2 \Theta)
\end{pmatrix}\, . \label{cC}
\ee
Equation~(\ref{cC})  agrees with \cite{Perotti:2018wxm}.

\begin{figure}[ht!]
\begin{center}
\includegraphics[width=2.5in]{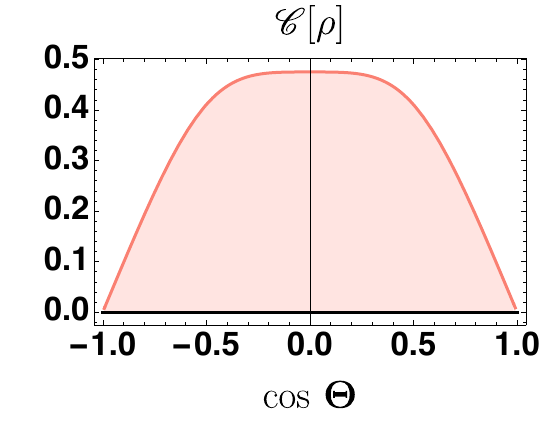}
\includegraphics[width=2.5in]{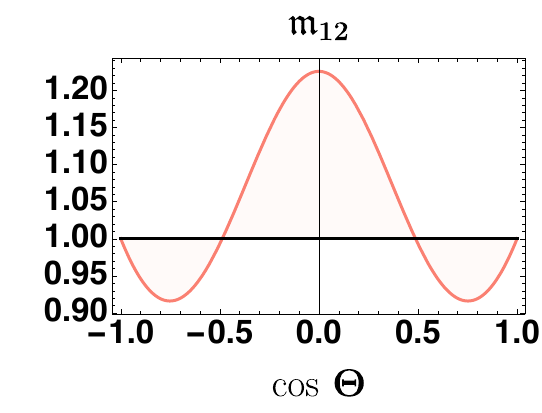}\vskip0.3cm
\includegraphics[width=2.5in]{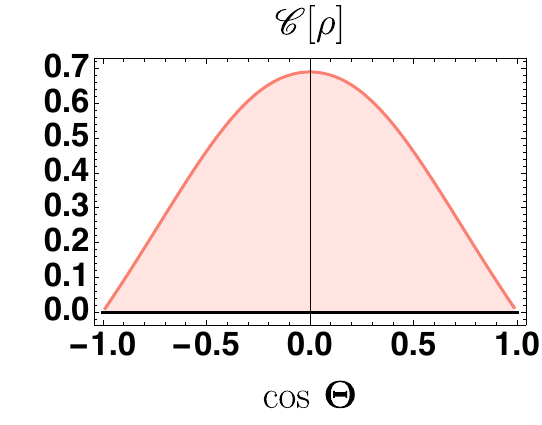}
\includegraphics[width=2.5in]{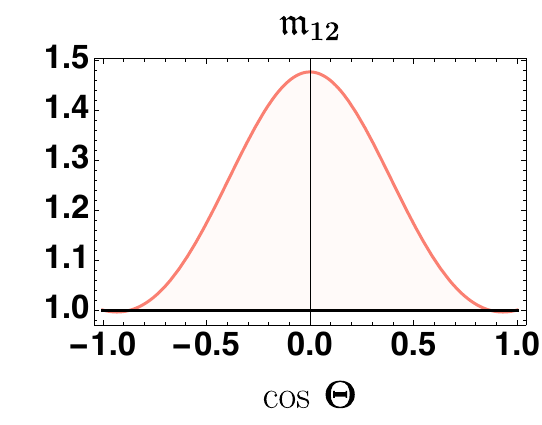}
\caption{\footnotesize In the first row:  Concurrence (left) and Horodecki condition $\mathfrak{m}_{12}$ (right) for $J/\psi \to \Lambda \xbar{\Lambda}$. In the bottom row: Concurrence (left) and Horodecki condition $\mathfrak{m}_{12}$ (right) for $\psi(3686) \to \Lambda \xbar{\Lambda}$.  All quantities are the largest for $\Theta=\pi/2$.
\label{fig:Jpsipsi} 
}
\end{center}
\end{figure}

As shown in Fig.~\ref{fig:Jpsipsi}, the concurrence and $\mathfrak{m}_{12}$ depend on the scattering angle.
The largest values are found at  $\Theta = \pi/2$, for which

\be
\mathscr{C}[\rho] =0.475 \pm 0.004\quad \text{and} \quad \mathfrak{m}_{12}  = 1.225 \pm 0.004\, ,
\ee
and the violation of the Bell inequality is established with a significance of $56.3\sigma$. 

 The same analysis can  be followed for the case of the $\psi(3686)$.   
 
 The decay $\psi(3686) \to \Lambda \bar \Lambda$ has branching fraction $(3.81\pm 0.13) \times 10^{-4}$~\cite{ParticleDataGroup:2022pth}. Events at energies around the value of the mass of the $\psi(3686)$ have been collected by selecting the $\Lambda$ and $\bar \Lambda$ decays into $p\pi^-$ and $\bar p\pi^-$ respectively. A likelihood fit yields  the helicity amplitude parameters $\alpha$ and $\Delta\Phi$~\cite{BESIII:2023euh}:
  \be
 \alpha= 0.69\pm 0.07 |_{\rm stat}\pm 0.02|_{\rm syst} \quad \text{and} \quad \Delta \Phi= 0.40^{+0.15}_{-0.14} |_{\rm stat} \pm 0.03 |_{\rm syst}\, .
 \ee 
No correlation between the uncertainties is given.

As shown in Fig.~\ref{fig:Jpsipsi}, the concurrence and $\mathfrak{m}_{12}$ are, as before, the largest  at  $\Theta = \pi/2$, for which
\be
\mathscr{C}[\rho] =0.69 \pm 0.07\quad \text{and} \quad \mathfrak{m}_{12}  = 1.48 \pm 0.10\, ,
\ee
and the violation of the Bell inequality is established with  a significance of $4.8\sigma$.

\subsection[$J/\psi \to \Xi^- + \bar \Xi^+\quad \text{and} \quad \Xi^0 + \bar \Xi^0$]{$\boxed{J/\psi \to \Xi^- + \bar \Xi^+ \quad \text{and} \quad \Xi^0 + \bar \Xi^0}$ }

We can just retrace our steps from the previous Section. 

The density and correlation matrices are the same as in \eq{rhoLambda} and \eq{cC}. The decay $J/\psi \to \Xi^- \bar \Xi^+$ has branching fraction
$(9.7\pm 0.8)  \times 10^{-4}$~\cite{ParticleDataGroup:2022pth}. The chain decays $ \Xi^- \to \Lambda \pi^-$, $\Lambda\to p\pi^-$ and  $\bar \Xi^+ \to \bar \Lambda \pi^+$,  $\bar \Lambda  \to \bar p \pi^+$ out of a sample of $1.31\times 10^{9}$ $J/\psi$ have been reconstructed. A maximum likelihood fit over the kinematic variables yields the values of the parameters defining the helicity amplitudes~\cite{BESIII:2021ypr}:
\be
\alpha=0.586 \pm 0.012 |_{\rm stat} \pm0.010 |_{\rm syst} \quad \text{and } \quad \Delta\Phi=1.213 \pm 0.046 |_{\rm stat} \pm 0.016 |_{\rm syst}\, .
\ee

\begin{figure}[ht!]
\begin{center}
\includegraphics[width=2.5in]{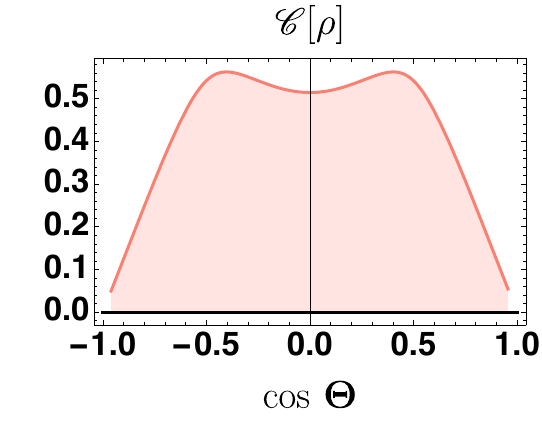}
\includegraphics[width=2.5in]{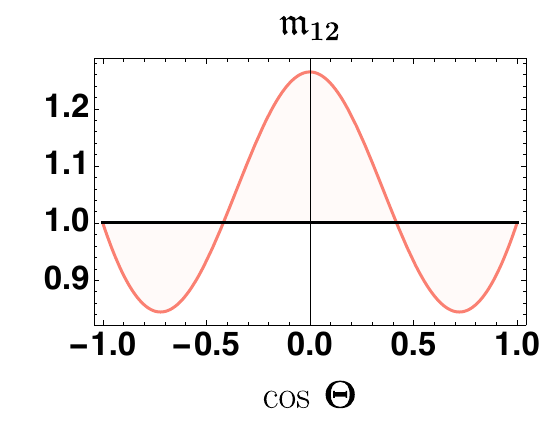}\vskip0.3cm
\includegraphics[width=2.5in]{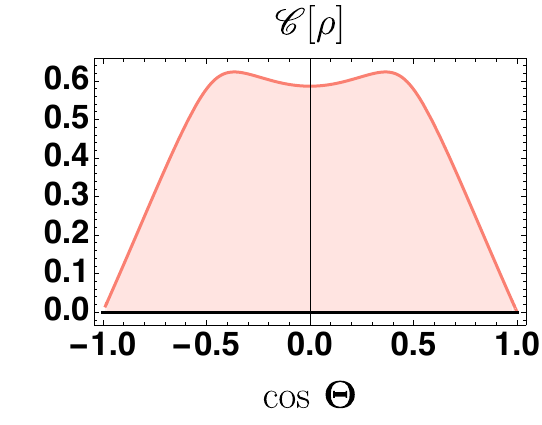}
\includegraphics[width=2.5in]{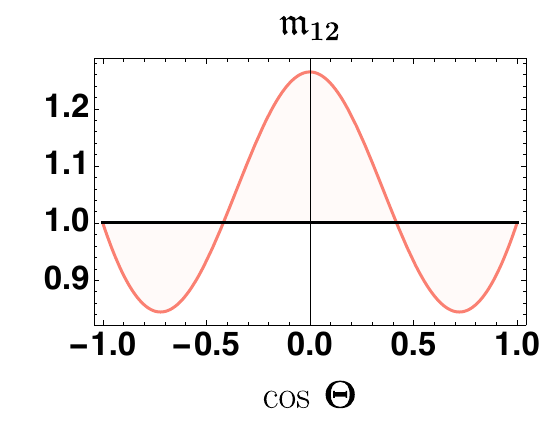}
\caption{\footnotesize In the first row:  Concurrence (left) and Horodecki condition $\mathfrak{m}_{12}$ (right) for $J/\psi \to \Xi^0\bar  \Xi^0$. In the bottom row: Concurrence (left) and Horodecki condition $\mathfrak{m}_{12}$ (right) for$J/\psi \to \Xi^+\bar \Xi^-$.   $\mathfrak{m}_{12}$  is the largest for $\Theta=\pi/2$.
. 
\label{fig:JpsiXi00pm} 
}
\end{center}
\end{figure}

As shown in Fig.~\ref{fig:JpsiXi00pm}, the concurrence and $\mathfrak{m}_{12}$ depend on the scattering angle.
The largest value for  $\mathfrak{m}_{12}$ is found at  $\Theta = \pi/2$, for which
\be
\mathscr{C}[\rho] =0.586\pm0.016  \quad \text{and} \quad \mathfrak{m}_{12}  = 1.343\pm0.018 \, .
\ee
This process exemplifies the fact that the largest violation of the Bell inequality does not necessary entail the largest value for the entanglement. The significance of the violation of the Bell inequality is $19.1\sigma$.

The next decay, $J/\psi \to \Xi^0 \bar \Xi^0$, has branching fraction $1.17 \times 10^{-3}$~\cite{BESIII:2021ypr}. A sample of 327 305 events is selected from the chain decays $\Xi^0 \to \Lambda \pi^0$, $\Lambda\to \to p\pi^-$ and  $ \bar \Xi^0 \to \bar \Lambda \pi^0$,  $\bar \Lambda  \to \bar p \pi^+$; the helicity amplitude parameters obtained from the likelihood fit are~\cite{BESIII:2023drj}:
\be
\alpha=0.514\pm 0.006 |_{\rm stat} \pm0.0015 |_{\rm syst} \quad \text{and } \quad \Delta\Phi=1.168\pm 0.019 |_{\rm stat} \pm 0.018 |_{\rm syst}\, .
\ee

As shown in Fig.~\ref{fig:JpsiXi00pm}, the concurrence and $\mathfrak{m}_{12}$ depend on the scattering angle. The largest value for  $\mathfrak{m}_{12}$ is found at  $\Theta = \pi/2$, for which
\be
\mathscr{C}[\rho] =0.514\pm 0.016\quad \text{and} \quad \mathfrak{m}_{12}  = 1.264\pm 0.017\,.
\ee
The significance of the violation of the Bell inequality is $15.5\sigma$.


\subsection[$\psi (3686)  \to\Xi^- + \bar \Xi^+ \quad \text{and} \quad \Xi^0 + \bar \Xi^0$]{$\boxed{\psi (3686)  \to\Xi^- + \bar \Xi^+ \quad \text{and} \quad \Xi^0 + \bar \Xi^0}$ }

In the process
\be
e^{+}e^{-} \to\gamma \to  c \bar c \to \psi (3686)  \to \Xi^- + \bar \Xi^+\, ,
\ee
the $\psi (3686)$ is produced polarized. The correlation matrix of the two baryons depends on the scattering angle $\Theta$ because the polarization of the $ \psi (3686)$ does. The process is similar to the previous ones and, again, we can simply retrace our steps from the previous Section. 

The density and correlation matrices are the same as in \eq{rhoLambda} and \eq{cC}. 

The decay $\psi(3686) \to \Xi^-  \bar \Xi^+$ has branching fraction $(2.87\pm 0.11) \times 10^{-4}$~\cite{ParticleDataGroup:2022pth}. The chain decays $\Xi^- \to \Lambda \pi^-$, $\Lambda\to p\pi^-$ and  $\bar \Xi^+ \to \bar \Lambda \pi^+$,  $\bar \Lambda  \to \bar p \pi^+$ have been reconstructed out of a sample of $(448.1\pm 2.9)\times 10^{6}$ $\psi(3686)$. A maximum likelihood fit over the kinematic variables yields the values of the parameters defining the helicity amplitudes~\cite{BESIII:2022lsz}:
  \be
 \alpha= 0.693 \pm 0.048 |_{\rm stat}\pm 0.049|_{\rm syst} \quad \text{and} \quad \Delta \Phi= 0.667\pm0.111|_{\rm stat} \pm 0.058|_{\rm syst}\,.
 \ee 
 
\begin{figure}[ht!]
\begin{center}
\includegraphics[width=2.5in]{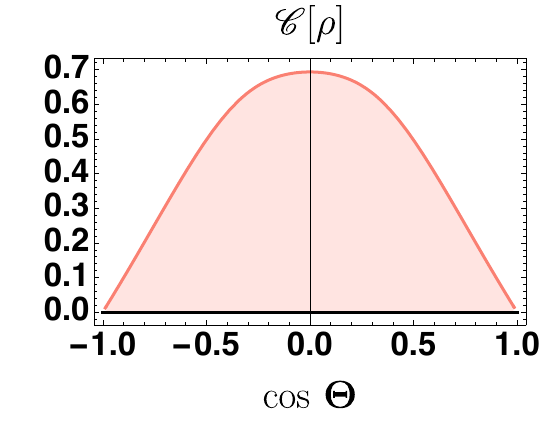}
\includegraphics[width=2.5in]{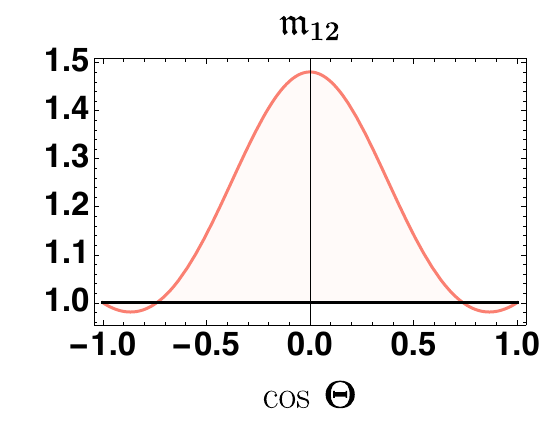}\vskip0.3cm
\includegraphics[width=2.5in]{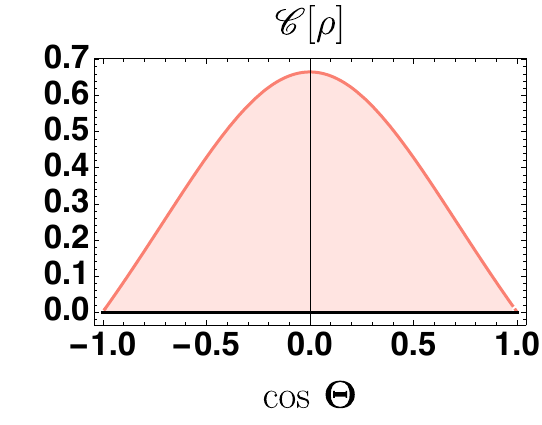}
\includegraphics[width=2.5in]{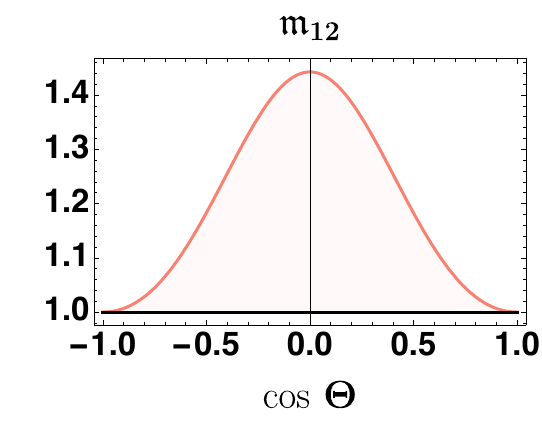}
\caption{\footnotesize  In the first row:  Concurrence (left) and Horodecki condition $\mathfrak{m}_{12}$ (right) for $\psi(3686) \to \Xi^0\bar \Xi^0$. In the bottom row:   Concurrence (left) and Horodecki condition $\mathfrak{m}_{12}$ (right) for $\psi(3686) \to \Xi^+\bar \Xi^-$. All quantities are the largest for $\Theta=\pi/2$.
\label{fig:psiXipm00pm} 
}
\end{center}
\end{figure}

We find 
\be
\mathscr{C}[\rho] =0.693\pm0.068 \quad \text{and} \quad \mathfrak{m}_{12}  = 1.480\pm 0.095\,,
\ee
at $\Theta =\pi/2$.
The significance of the violation of the Bell inequality is $5.1\sigma$.

The next decay, $\psi(3686) \to \Xi^0  \bar \Xi^0$, has branching fraction $(2.3 \pm 0.4)\times 10^{-4}$~\cite{ParticleDataGroup:2022pth}. The chain decays $J/\psi \to \Xi^0 \to \Lambda \pi^0$, $\Lambda \to p\pi^-$ and  $J/\psi \to \bar \Xi^0 \to \bar \Lambda \pi^0$,  $\bar \Lambda  \to \bar p \pi^+$ have been reconstructed out of the same sample of $\psi(3686)$ as before; the helicity amplitude parameters obtained from the likelihood fit are~\cite{BESIII:2023lkg}: 
   \be
 \alpha= 0.665 \pm 0.086 |_{\rm stat}\pm 0.081|_{\rm syst} \quad \text{and} \quad \Delta \Phi= -0.050\pm0.150|_{\rm stat} \pm 0.020|_{\rm syst}\, .
 \ee 

We find 
\be
\mathscr{C}[\rho] =0.665\pm 0.119\quad \text{and} \quad \mathfrak{m}_{12}  = 1.442\pm 0.161\, ,
\ee
at $\Theta =\pi/2$.

The significance of the violation of the Bell inequality is $2.7\sigma$.

\subsection[$J/\psi \to \Sigma^- + \bar \Sigma^+ \quad \text{and} \quad \psi(3686) \to \Sigma^- + \bar  \Sigma^+$]{$\boxed{J/\psi \to \Sigma^- + \bar \Sigma^+ \quad \text{and} \quad \psi(3686) \to \Sigma^- + \bar  \Sigma^+}$}

 In these decays, the correlation matrix of the two baryons depends on the scattering angle $\Theta$ because the polarization of the $ \psi (3686)$ does. Again, we can just retrace our steps from the previous Sections. The density and correlation matrices are the same as in \eq{rhoLambda} and \eq{cC}. 

The decay $J/\psi \to \Sigma^-  \bar \Sigma^+$ has branching fraction $(1.50\pm0.24) \times 10^{-3}$~\cite{ParticleDataGroup:2022pth}. The decays $\Sigma^- \to  \bar p\pi^0$ and  $\bar \Sigma^+ \to  p\pi^0$ out of a sample of $1.31\times 10^{9}$ $J/\psi$ have been reconstructed. A maximum likelihood fit over the kinematic variables yields the values of the parameters defining the helicity amplitudes~\cite{BESIII:2020fqg}:
\be
\alpha = -0.508 \pm 0.006 |_{\rm stat}\pm 0.004|_{\rm syst} \quad \text{and} \quad \Delta \Phi= -0.270\pm0.012|_{\rm stat} \pm 0.009|_{\rm syst} \, . 
\ee

\begin{figure}[ht!]
\begin{center}
\includegraphics[width=2.5in]{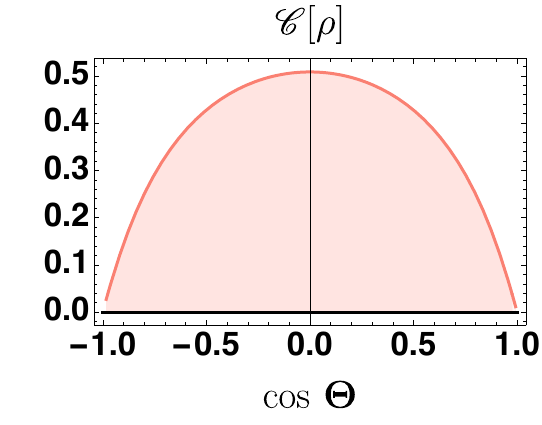}
\includegraphics[width=2.5in]{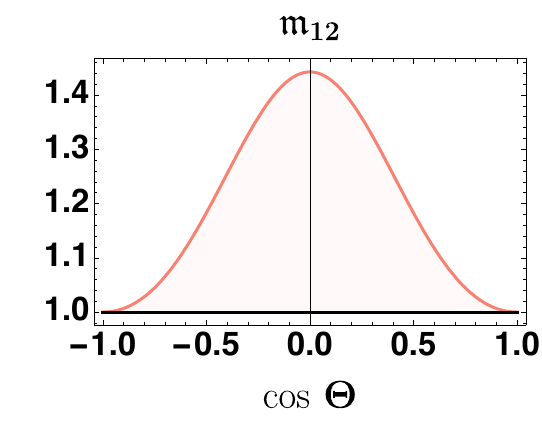}\vskip0.3cm
\includegraphics[width=2.5in]{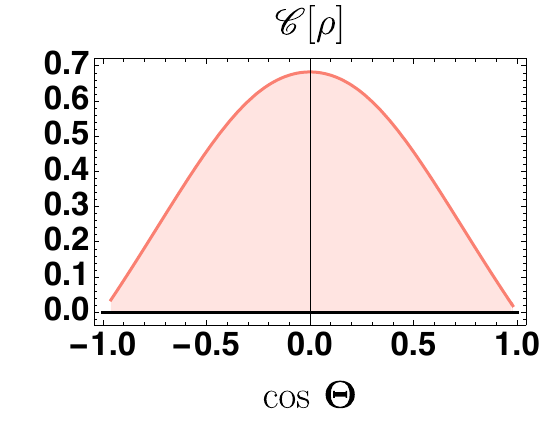}
\includegraphics[width=2.5in]{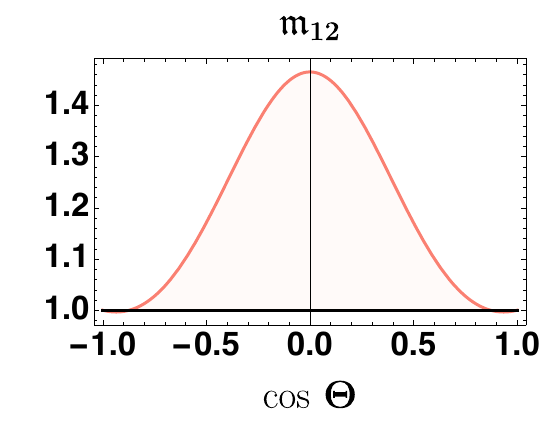}
\caption{\footnotesize In the first row:  Concurrence (left) and Horodecki condition $\mathfrak{m}_{12}$ (right) for $J/\psi \to \bar \Sigma^+ \Sigma^-$. In the bottom row:   Concurrence (left) and Horodecki condition $\mathfrak{m}_{12}$ (right) for $\psi(3686) \to \bar \Sigma^+ \Sigma^-$. All quantities are the largest for $\Theta=\pi/2$. All quantities are the largest for $\Theta=\pi/2$.
\label{fig:JpsipsiSigmapm} 
}
\end{center}
\end{figure}

We find 
\be
\mathscr{C}[\rho] =0.508 \pm 0.007\quad \text{and} \quad \mathfrak{m}_{12}  = 1.258\pm 0.007\, ,
\ee
at $\Theta =\pi/2$.

The significance of the violation of the Bell inequality is therefore 36.9$\sigma$.

The next decay, $\psi(3686) \to \Sigma^-  \bar \Sigma^+$, has branching fraction $(2.82\pm 0.09) \times 10^{-4}$~\cite{ParticleDataGroup:2022pth}. The decays $ \Sigma^- \to \bar  p\pi^0$ and  $ \bar \Sigma^+ \to  p \pi^+$ out of a sample of $4.48\times 10^{8}$ $\psi(3686) $ have been reconstructed. A maximum likelihood fit over the kinematic variables yields the values of the parameters defining the helicity amplitudes~\cite{BESIII:2020fqg}:
\be
\alpha = 0.682 \pm 0.030 |_{\rm stat}\pm 0.011|_{\rm syst} \quad \text{and} \quad \Delta \Phi= 0.379\pm0.07|_{\rm stat} \pm 0.014|_{\rm syst} \, .
\ee

We find 
\be
\mathscr{C} =0.682\pm 0.032 \quad \text{and} \quad \mathfrak{m}_{12}  = 1.465 \pm 0.043\, ,
\ee
at $\Theta =\pi/2$.

The significance of the violation of the Bell inequality is therefore 10.8$\sigma$.

\subsection[$J/\psi \to \Sigma^0 + \bar \Sigma^0 \quad \text{and} \quad \psi(3686) \to \Sigma^0 + \bar \Sigma^0 $]{$\boxed{J/\psi \to \Sigma^0 + \bar \Sigma^0 \quad \text{and} \quad \psi(3686) \to \Sigma^0 + \bar \Sigma^0 }$ }

\begin{figure}[ht!]
\begin{center}
\includegraphics[width=2.5in]{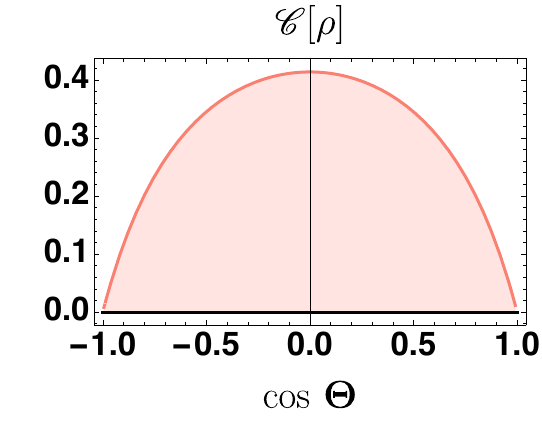}
\includegraphics[width=2.5in]{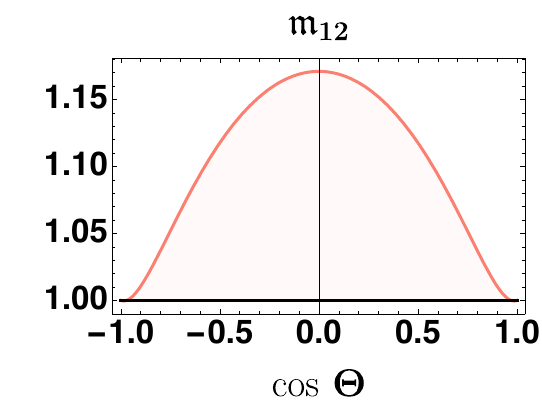}\vskip0.3cm
\includegraphics[width=2.5in]{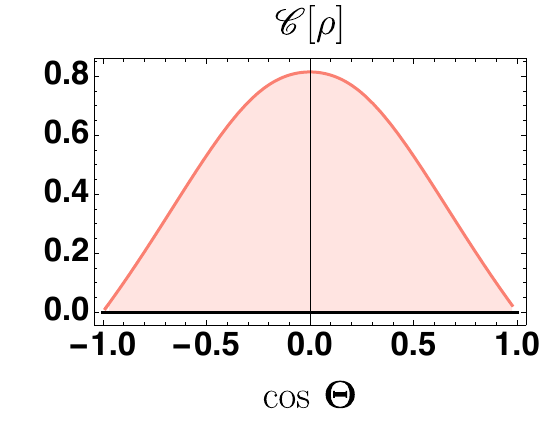}
\includegraphics[width=2.5in]{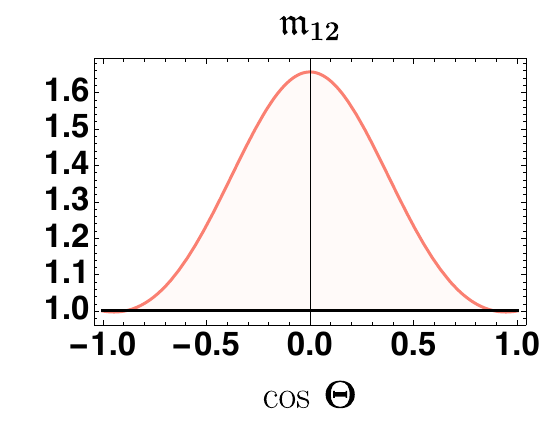}
\caption{\footnotesize In the first row:  Concurrence (left) and Horodecki condition $\mathfrak{m}_{12}$ (right) for $J/\psi \to \Sigma^0\bar \Sigma^0$. In the bottom row:   Concurrence (left) and Horodecki condition $\mathfrak{m}_{12}$ (right) for $\psi(3686) \to \Sigma^0\bar \Sigma^0$. All quantities are the largest for $\Theta=\pi/2$. All quantities are the largest for $\Theta=\pi/2$.
\label{fig:JpsipsiSigma00} 
}
\end{center}
\end{figure}

  One last time, we can just retrace our steps from the previous Sections. The density and correlation matrices are the same as in \eq{rhoLambda} and \eq{cC}. 

The decay $J/\psi  \to \Sigma^{0} \bar \Sigma^{0}$ has branching fraction $(1.172\pm0.032) \times 10^{-3}$~\cite{ParticleDataGroup:2022pth}. The chain decays $ \Sigma^0 \to  \Lambda \gamma$, $\Lambda \to p \pi^-$ and  $ \bar \Sigma^+ \to  \bar \Lambda \gamma$, $\bar \Lambda \to \bar p \pi^+$ have been reconstructed out of a sample of $1.0\times 10^{10}$ $J/\psi$. A maximum likelihood fit over the kinematic variables yields the values of the parameters defining the helicity amplitudes~\cite{BESIII:2024nif}:
\be
\alpha = -0.4133 \pm 0.0035 |_{\rm stat}\pm 0.0077|_{\rm syst} \quad \text{and} \quad \Delta \Phi= -0.0828\pm0.00068|_{\rm stat} \pm 0.0033|_{\rm syst} \,.
\ee

We find for the $J/\psi  \to \Sigma^{0} \bar \Sigma^{0}$
\be
\mathscr{C}[\rho]=0.4133\pm 0.0086 \quad \text{and} \quad \mathfrak{m}_{12}  = 1.171\pm 0.007\, ,
\ee
at $\Theta =\pi/2$.

The significance of the violation of the Bell inequality is  24.4$\sigma$.

The next decay, $\psi(3686) \to \Sigma^0  \bar \Sigma^0$, has branching fraction $(2.35\pm0.09) \times 10^{-4}$~\cite{ParticleDataGroup:2022pth}. The chain decays $ \Sigma^0 \to  \Lambda \gamma$, $\Lambda \to p \pi^-$ and  $ \bar \Sigma^+ \to  \bar \Lambda \gamma$, $\bar \Lambda \to \bar p\pi^+$ out of a sample of $2.7 \times 10^{9}$$\psi(3686) $ have been reconstructed. A maximum likelihood fit over the kinematic variables yields the values of the parameters defining the helicity amplitudes~\cite{BESIII:2024nif}:
\be
\alpha = 0.814 \pm 0.028 |_{\rm stat}\pm 0.028|_{\rm syst} \quad \text{and} \quad \Delta \Phi= 0.512\pm0.085|_{\rm stat} \pm 0.034|_{\rm syst} \, .
\ee

We find 
\be
\mathscr{C} [\rho]=0.814 \pm 0.040\quad \text{and} \quad \mathfrak{m}_{12}  = 1.663\pm0.065 \, ,
\ee
at $\Theta =\pi/2$.
The significance of the violation of the Bell inequality is therefore 10.2$\sigma$. 

\subsection{Entanglement as a function of the form factors}

\begin{figure}[ht!]
\begin{center}
\includegraphics[width=3.5in]{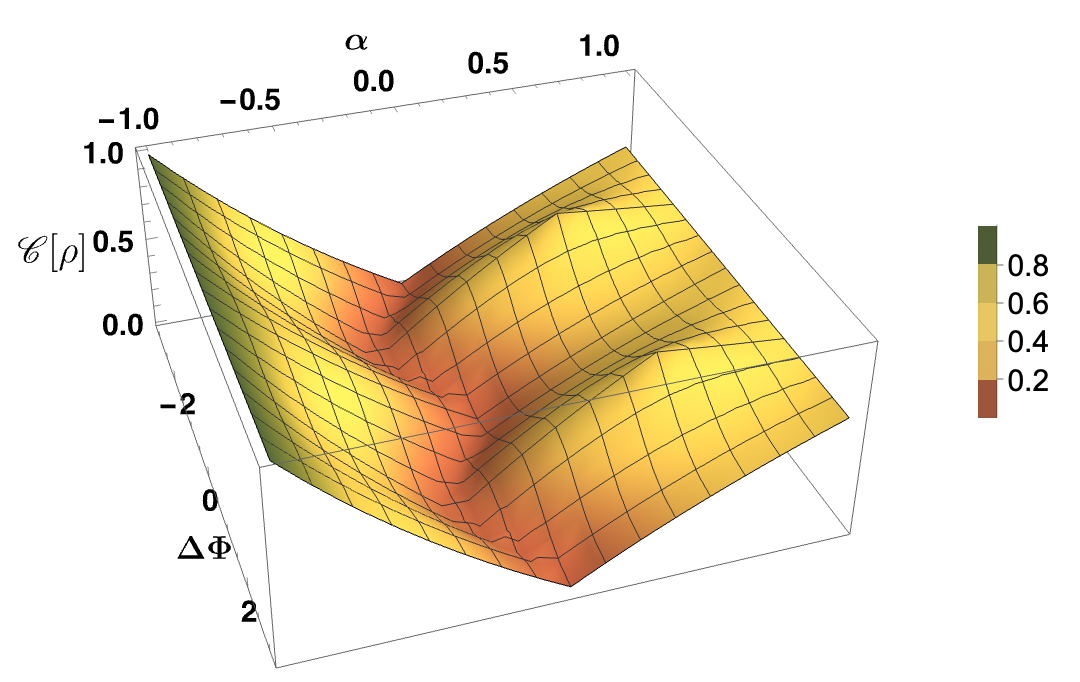}
\includegraphics[width=3in]{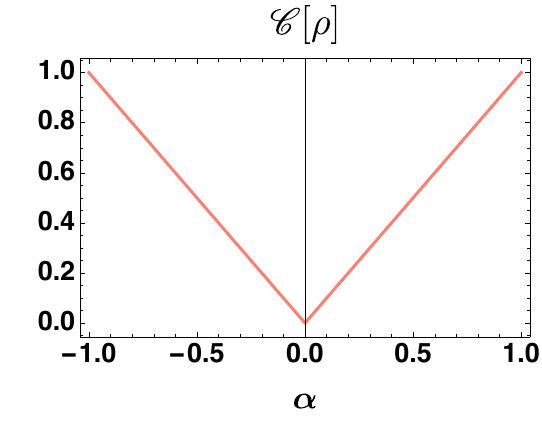}
\caption{\footnotesize Left side: Entanglement as a function of the form factor parameters $\alpha$ and $\Delta\Phi$. The concurrence is computed at $\Theta=\pi/3$. Right side: The concurrence as a function of the parameter $\alpha$ at $\Theta=\pi/2$ is $\mathscr{C} [\rho]= |\alpha|$. There is no dependence on the other parameter $\Delta\Phi$. 
\label{fig:form} 
} 
\end{center}
\end{figure}

We have seen that the amount of entanglement in the final state spin correlations depends on the kind of baryons and on the charmonium state these come from.

\begin{table}[h!]
\tablestyle[sansboldbw]
\begin{tabular}{*{3}{p{0.2\textwidth}}}
\theadstart
    \thead  decay &\thead $\mathfrak{m}_{12}$ &
    \thead significance \\
\tbody
 $J/\psi \to \Lambda \bar \Lambda$ & 1.225 $\pm$ 0.004& \hskip1cm 56.3 \\
 $\psi(3686) \to \Lambda \bar \Lambda$  &1.476 $\pm$ 0.100   & \hskip1cm 4.8 \\
 $J/\psi \to \Xi^- \bar \Xi^+$    &1.343 $\pm$ 0.018  &  \hskip1cm 19.1 \\
 $J/\psi \to \Xi^0 \bar \Xi^0$    &1.264  $\pm$ 0.017  &  \hskip1cm 15.6 \\
 $\psi(3686) \to\Xi^- \bar \Xi^+$  &1.480 $\pm$ 0.095  & \hskip1cm 5.1\\
$\psi(3686) \to\Xi^0 \bar \Xi^0$  &1.442 $\pm$ 0.161 &  \hskip1cm 2.7 \\
 $J/\psi \to \Sigma^- \bar \Sigma^+$    &1.258 $\pm$ 0.007 & \hskip1cm 36.9 \\
 $\psi(3686) \to\Sigma^- \bar \Sigma^+$  &1.465 $\pm$ 0.043  &  \hskip1cm 10.8 \\
  $J/\psi \to \Sigma^0 \bar \Sigma^0$   &1.171 $\pm$ 0.007   & \hskip1cm 24.4\\
$\psi(3686) \to\Sigma^0 \bar \Sigma^0$  &1.663 $\pm$ 0.065   & \hskip1cm 10.2 \\
   \hline%
  \tend
\end{tabular}
\caption{\footnotesize \label{tab:fermions} \textrm{Summary of Bell inequality violation in spin 1 charmonium decays into baryons.}}
\end{table}

 We can study how in general the entanglement varies as we vary the form factors $G_{M}$ and $G_{E}$ in the coupling of the baryons to the spin 1 resonance. For instance, the coupling of the $\Lambda$ baryons to the $J/\psi$ is given as
\be
\bar u_{\Lambda} \left[ F_{1} \gamma^{\mu} + \frac{1}{2 m_{\Lambda}} \sigma^{\mu\nu} q^{\nu} F_{2}\right] u_{\Lambda} A_{\mu}^{J/\psi}\label{current}
\ee
in which $G_{M}=F_{1}+F_{2}$ and $G_{E}= F_{1}+s/2 m_{\Lambda}^{2} F_{2}$ and $s=q^2$ is the square of the energy. The field $ A_{\mu}^{J/\psi}$ represents the $J/\psi$ charmonium state. In \eq{current}, $\sigma^0=\mathbb{1}$ and $\sigma^{\mu\nu}=i/2 \{ \sigma^\mu,\, \sigma^\nu \}$.

The two form factors  $G_{M}$ and $G_{E}$ have been written  in terms of the two coefficients $\alpha$ and $\Delta \Phi$ as
\be
\frac{G_{M}}{G_{E} }= \left|\frac{G_{M}}{G_{E}} \right|\, e^{i \Delta \Phi}\quad \text{and} \quad 
\alpha=  \frac{s |G_{M}|^{2}-4m_{\Lambda}^{2}|G_{E}|^{2} }{s |G_{M}|^{2}+4m_{\Lambda}^{2}|G_{E}|^{2}} \, .
\ee

The left-hand side of Fig.~\ref{fig:form} shows the variation of the  entanglement as we vary these two parameters. It is the largest when $\alpha=-1$, that is when $G_{M}=0$. 
This value corresponds to  $F_{1}=-F_{2}$, which gives the coupling of an elementary scalar to the photon and is parity conserving. The other limit of interest is $F_{2}=0$, for which  $G_{E}=G_{M}$ and corresponds at threshold to  $\alpha=1$. This is the minimal coupling of an elementary fermion to the photon.

Setting the scattering angle to $\Theta=\pi/2$, the concurrence ceases to depend on the phase $\Delta\Phi$---as it can be seen by inspection of \eq{rhoLambda}---and becomes identical to the absolute value of the coefficient $\alpha$, see the right-hand side of Fig.~\ref{fig:form}

The values of entanglement and Bell inequality violation for the decays of charmonium states with spin 1 into baryons are summarized in Table~\ref{tab:fermions}.

\subsection[$\chi^{1}_{c}\to\phi +\phi$]{$\boxed{\chi^{1}_{c}\to\phi +\phi}$}

Though more complicated, the computation for this process is essentially  similar to the previous one. 

The $\chi^{1}_c$ are produced in
\be
e^{+} e_{-} \to \psi (3686) \to \gamma \chi^{1}_c\,,
\ee
with a branching fraction of $(4.36 \pm 0.13 \pm 0.18) \times 10^{-4}$~\cite{BESIII:2023zcs}. The elements of the spin density matrix can be written as 
\be
\rho_{\lambda_{1}\lambda_{2},\lambda_{1}^{\prime}\lambda_{2}^{\prime}}= w_{\lambda_{1}\lambda_{2}}w^{*}_{\lambda_{1}^{\prime}\lambda_{2}^{\prime}} \sum_{k=\pm1} D^{(1)*}_{k,\lambda_{1}-\lambda_{2}}(0,\Theta,0) D^{(1)}_{k,\lambda_{1}^{\prime}-\lambda_{2}^{\prime}}(0,\Theta,0)
\ee
where $D^{(1)}_{i,j}$ is the Wigner matrix for the spin 1 representation of $SO(3)$.
\begin{figure}[ht!]
\begin{center}
\includegraphics[width=3.5in]{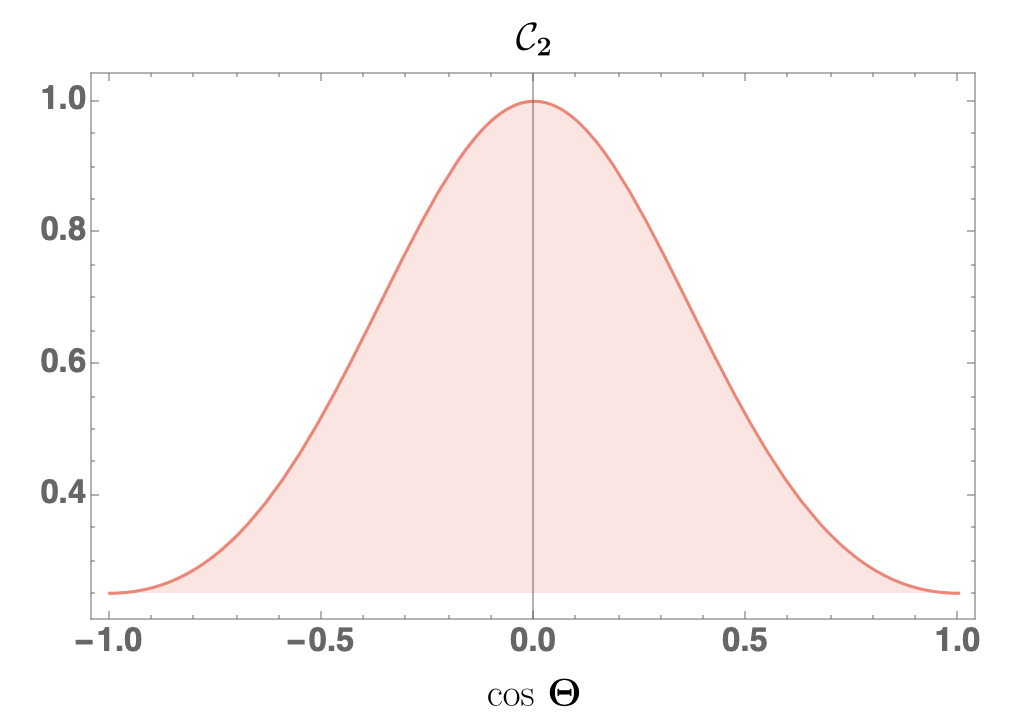}
\caption{\footnotesize Dependence of the operator $\cmb$ on the scattering angle $\Theta$ for  $\chi_c^1 \to \phi \phi$.
\label{fig:c2phiphi} 
}
\end{center}
\end{figure}

There are 2 independent amplitudes, because the symmetry under exchange of identical particles in the final state gives $ w_{_{1\,1}} =  w_{_{0,0}} = 0$. The density matrix is given by
{\small
\be 
\rho_{\phi\phi} \propto \left(
\begin{array}{ccccccccc}
 0 & 0 & 0 & 0 & 0 & 0 & 0 & 0 & 0 \\
 0 &  |w_{-1\,0}|^2(2- s^2_{\Theta})  & 0 &
    w_{-1\,0} w^{\star}_{0\,-1} s^2_{\Theta} & 0 &  w_{-1\,0}
   w^{\star}_{0\,1} (2-s^2_{\Theta})  & 0 &  w_{-1\,0}
   w^{\star}_{1\,0} s^2_{\Theta} & 0 \\
 0 & 0 & 0 & 0 & 0 & 0 & 0 & 0 & 0 \\
 0 &  w_{0\,-1} w^{\star}_{-1\,0} s^2_{\Theta} & 0 &
 |w_{0\,-1}|^2(2-s^2_{\Theta}) & 0 &  w_{0\,-1}
   w^{\star}_{0\,1} s^2_{\Theta} & 0 &  w_{0\,-1} w_{1\,0}(2-s^2_{\Theta})
    & 0 \\
 0 & 0 & 0 & 0 & 0 & 0 & 0 & 0 & 0 \\
 0 &  w_{0\,1} w^{\star}_{-1\,0} (2-s^2_{\Theta}) & 0 &
 w_{0\,1} w^{\star}_{0\,-1} s^2_{\Theta} & 0 &
 |w_{0\,1}|^2(2-s^2_{\Theta}) & 0 &  w_{0\,1}
   w^{\star}_{1\,0} s^2_{\Theta} & 0 \\
 0 & 0 & 0 & 0 & 0 & 0 & 0 & 0 & 0 \\
 0 &  w_{1\,0} w^{\star}_{-1\,0} s^2_{\Theta} & 0 &  w_{1\,0}
   w^{\star}_{0\,-1}  (2-s^2_{\Theta}) & 0 &  w_{1\,0}
   w^{\star}_{0\,1} s^2_{\Theta} & 0 &  |w_{1\,0}|^2(2-s^2_{\Theta})
    & 0 \\
 0 & 0 & 0 & 0 & 0 & 0 & 0 & 0 & 0 \\
\end{array}
\right)\, ,
\ee
}
in which $s_{\Theta}\equiv \sin{\Theta}$.

The analysis of the data in~\cite{BESIII:2023zcs}  selects $1529\pm45$ out of the $\gamma K^+K^-K^+K^-$ final state events. The maximum likelihood fit yields the absolute value of the ratio of the moduli of the helicity amplitudes:
\be
 \left| \frac{w_{_{1\,0}}}{ w_{_{0,1}}}\right|  = 1.05\pm0.05 \quad \text{and} \quad
  \left| \frac{w_{_{1\,1}}}{ w_{_{1\,0}}} \right|= 0.07 \pm 0.04\, ,
 \ee
in which the uncertainty is only  statistical. The symmetries of the decay process imply that $ w_{_{0\,1}} = - w_{_{-1,0}} $ and $w_{_{1\,0}} =  -w_{_{0\,-1}}$.

There is no easily computable direct measure of the entanglement for a bipartite system of two qutrits. We  compute instead the quantity $\cmb$, defined in \eq{eq:cmb-qutrit}, which is a lower bound on it. It is maximum at scattering angle $\Theta=\pi/2$, as shown in  Fig.~\ref{fig:c2phiphi}, where it is 1, showing that also the entanglement must be maximal.

Concerning the violation of the Bell inequality, we find (at $\Theta=\pi/2$)
\be
\Tr \mathscr{B}  \rho_{\phi\phi} = 2.296 \pm 0.003\, .
\ee
The significance of the violation of the Bell inequality is $98.7\sigma$---though we must bear in mind that the uncertainties used in this estimate are only statistical.

The $\phi$ are polarized and their polarization is   given by
\be
 \Tr \rho_{\phi\phi}  \, S_{i}  \otimes \mathbb{1} = (0,\,0,\, 0.024)\, ,
\ee
where $S_{1}= \dfrac{1}{\sqrt{2}} \left( T^{1}+T^{6} \right)$, $S_{2}= \dfrac{1}{\sqrt{2}} \left( T^{2}+T^{7} \right)$ and $S_{3}= \dfrac{1}{2} T^{3}+ \dfrac{\sqrt{3}}{2} T^{8}$.

\subsection[$\psi(3686) \to \Omega^{-} + \bar \Omega^{+}$]{$\boxed{\psi(3686) \to \Omega^{-} + \bar \Omega^{+}}$ }

The elements of the $16\times16$ density matrix describing the final state can be written as 
\be
\rho_{\lambda_{1}\lambda_{2},\lambda_{1}^{\prime}\lambda_{2}^{\prime}}\propto w_{\lambda_{1}\lambda_{2}}w^{*}_{\lambda_{1}^{\prime}\lambda_{2}^{\prime}} \sum_{k=\pm1} D^{(1)*}_{k,\lambda_{1}-\lambda_{2}}(0,\Theta,0) D^{(1)}_{k,\lambda_{1}^{\prime}-\lambda_{2}^{\prime}}(0,\Theta,0)
\ee
where $D^{(1)}_{i,j}$ is the Wigner matrix for the spin 1 representation of $SO(3)$.

There are 16 helicity amplitudes but 8 vanish because of the condition $|\lambda_1-\lambda_2| \leq1$ and the symmetry of the final state. The remaining are related by parity and by the symmetries of the final state as:
\bea
 w_{_{-\frac{3}{2}\,-\frac{3}{2}} } &= & - w_{\frac{3}{2}\,\frac{3}{2}} \\
   w_{-\frac{1}{2}\,-\frac{1}{2}} &=&  - w_{\frac{1}{2}\,\frac{1}{2}}\\
  w_{_{-\frac{1}{2}\,\frac{1}{2}} } &= &w_{\frac{3}{2}\,-\frac{1}{2}}\\
  w_{_{-\frac{1}{2}\,-\frac{3}{2}} } = w_{-\frac{3}{2}\,-\frac{1}{2}} &=&
 - w_{\frac{1}{2}\,\frac{3}{2}} =   -w_{\frac{3}{2}\,\frac{1}{2}} \, .
\eea
There are therefore  only four independent amplitudes: $w_{\frac{3}{2}\, \frac{3}{2}}$, $  w_{\frac{1}{2}\, \frac{1}{2}}$, $w_{\frac{1}{2}\, -\frac{1}{2}}$ and $w_{\frac{3}{2}\, \frac{1}{2}}$.

A selection of 4000 events with $\Omega^-\bar \Omega^+$ in the final state is taken out of a sample of $(448.1\pm2.9)\times 10^{6}$  $\psi(3686)$  collected~\cite{BESIII:2020lkm}. The decay chain is $\psi(3686)\to\Omega^-\bar \Omega^+$ and $\Omega^- \to K^-\Lambda (\to p \pi^-)$ and $\bar \Omega^+\to K^+\Lambda (\to \bar p \pi^+)$. 
  
The values of the four amplitudes are given  in the parametrization
  \be
  \frac{w_{\frac{1}{2}\, \frac{1}{2}}}{w_{\frac{1}{2}\, -\frac{1}{2}}} = h_1 e^{i \phi_1} \, , \quad
  \frac{w_{\frac{3}{2}\, \frac{1}{2}}}{w_{\frac{1}{2}\, -\frac{1}{2}}} = h_3 e^{i \phi_3}\,, \quad \text{and} \quad
  \frac{w_{\frac{3}{2}\, \frac{3}{2}}}{w_{\frac{1}{2}\, -\frac{1}{2}}} = h_4 e^{i \phi_4}
  \ee
 with two possible sets of solutions~\cite{BESIII:2020lkm}: 
 \bea
 h_1=0.30 \pm 0.11 |_{\rm stat}\pm 0.04|_{\rm syst}\quad \phi_1=0.69\pm 0.041 |_{\rm stat}\pm 0.13|_{\rm syst}\\
  h_3=0.26\pm 0.05 |_{\rm stat}\pm 0.02|_{\rm syst}\quad \phi_3=2.60\pm 0.16 |_{\rm stat}\pm 0.08|_{\rm syst}\\
   h_4 = 0.51 \pm 0.03 |_{\rm stat}\pm 0.01|_{\rm syst}\quad \phi_4=0.34\pm 0.80 |_{\rm stat}\pm 0.31|_{\rm syst}
   \eea
   and
   \bea
 h_1=0.31  \pm 0.10|_{\rm stat}\pm 0.04|_{\rm syst}\quad \phi_1=2.38 \pm 0.37|_{\rm stat}\pm 0.13|_{\rm syst}\\
  h_3=0.27\pm 0.05|_{\rm stat}\pm 0.01|_{\rm syst} \quad \phi_3=2.57\pm 0.16|_{\rm stat}\pm 0.04|_{\rm syst} \\
   h_4 = 0.51\pm 0.03|_{\rm stat}\pm 0.01|_{\rm syst} \quad \phi_4=1.37\pm 0.68|_{\rm stat}\pm 0.16|_{\rm syst}
 \eea

We find a negativity (at $\Theta=\pi/2$) \
\be
{\cal N}(\rho) =0.71 \pm 0.04\; \text{(sol I)}\quad \text{and} \quad  {\cal N}(\rho) = 1.34\pm0.03 \; \text{(sol II)}\, ,
\ee
 and therefore a substantial amount of entanglement. 
Although Bell inequalities involving particle of any spin have been discussed in the literature,
a reliable estimator to test Bell inequality between spin 3/2 fermions is not yet available.

\section{Charmonium spin 2 state}
\subsection[$\chi^2_c \to\phi \phi$]{$\boxed{\chi^2_c \to\phi \phi}$}

The $\chi^{2}_{c}$ are produced in
\be
e^{+} e^{-} \to \psi (3686) \to \gamma \chi^{2}_{c}\,,
\ee

The elements of the  density matrix can be written as 
\be
\rho_{\lambda_{1}\lambda_{2},\lambda_{1}^{\prime}\lambda_{2}^{\prime}}\propto w_{\lambda_{1}\lambda_{2}}w^{*}_{\lambda_{1}^{\prime}\lambda_{2}^{\prime}} \sum_{k=\pm1 \pm2 } D^{(2)*}_{k,\lambda_{1}-\lambda_{2}}(0,\Theta,0) D^{(2)}_{k,\lambda_{1}^{\prime}-\lambda_{2}^{\prime}}(0,\Theta,0)
\ee
where $D^{(2)}_{i,j}$ is the Wigner matrix for the spin 2 representation of $SO(3)$ and the sum is only over the polarization $\pm1$ and $\pm2$ because the spin 2 state is produced from unpolarized electrons and positrons with the electron and positron taken to be massless and, therefore, with only the helicities $\pm1$ and $\pm2$.

The analysis of the data in~\cite{BESIII:2023zcs}  selects $4247\pm93$ out of the $\gamma K^+K^-K^+K^-$ final state events. The maximum likelihood fit yields the absolute value of the ratio of the moduli of the helicity amplitudes: 
\bea
\left| \frac{ w_{_{1\,1}} }{ w_{_{0\,0}} }\right|=&=& 0.808\pm 0.051|_{\rm stat}\pm 0.009|_{\rm syst}  \\
\left| \frac{ w_{_{1\,-1}} }{ w_{_{0\,0}} }\right|=&=& 1.450\pm 0.097|_{\rm stat}\pm 0.104|_{\rm syst}  \\
 \left| \frac{ w_{_{0\,1}} }{ w_{_{0\,0}} }\right|=&=& 1.265\pm 0.054|_{\rm stat}\pm 0.079|_{\rm syst}  \,.
 \eea
Only the moduli of the amplitudes are given. The other amplitudes are related to these as follows
 \be
 w_{_{-1\,-1}} =w_{_{1\,1}} \, ,\quad  w_{_{-1\,1}} = w_{_{1\,-1}}\, , \quad \text{and}  \quad w_{_{0\,-1}}= w_{_{1\,0}} =w_{_{-1\,0}}= w_{_{0\,1}}\, .
 \ee
We find no indication of entanglement, namely 
\be
{\cal N} (\rho) = 0\, , \quad \text{and} \quad  \cmb =0\, . 
\ee
The expectation value of the Bell operator is given, around $\Theta =\pi/2$, by
\be
\Tr \mathscr{B}  \rho_{\phi\phi} =  1.202 \pm 0.032  
\ee
and the Bell inequality is not  violated.

The $\phi$ are not polarized:
\be
\Tr \rho_{\phi\phi} \, S_{i}  \otimes \mathbb{1} = (0,\, 0,\,  0) \, .
\ee

\section{Quantum correlation and decoherence}

{\versal  The energy available }to the $\Lambda$ baryons in, for instance, the decay $\eta_c \to \Lambda \bar \Lambda$  makes these particles travel at about $0.66\, c$ and decay, in average, around 7  centimeters away from the primary vertex.
The beam pipe at BESIII has an inner (outer) diameter of 6.3 (11.4)  cm~\cite{BESIII:2009fln} while  at the LHC$b$ it goes from  6.5 (near the interaction region) to 26.2 cm with a conical design~\cite{LHCb:2008vvz}.  It therefore seems very possible that the $\Lambda$ baryons  do hit the wall of the beam pipe and even go inside the first layers of the detector, which is the multilayer drift chamber (MDC) at BESIII and the vertex detector (VELO) at LHC$b$. 
\begin{figure}[h!]
\begin{center}
\includegraphics[width=3.5in]{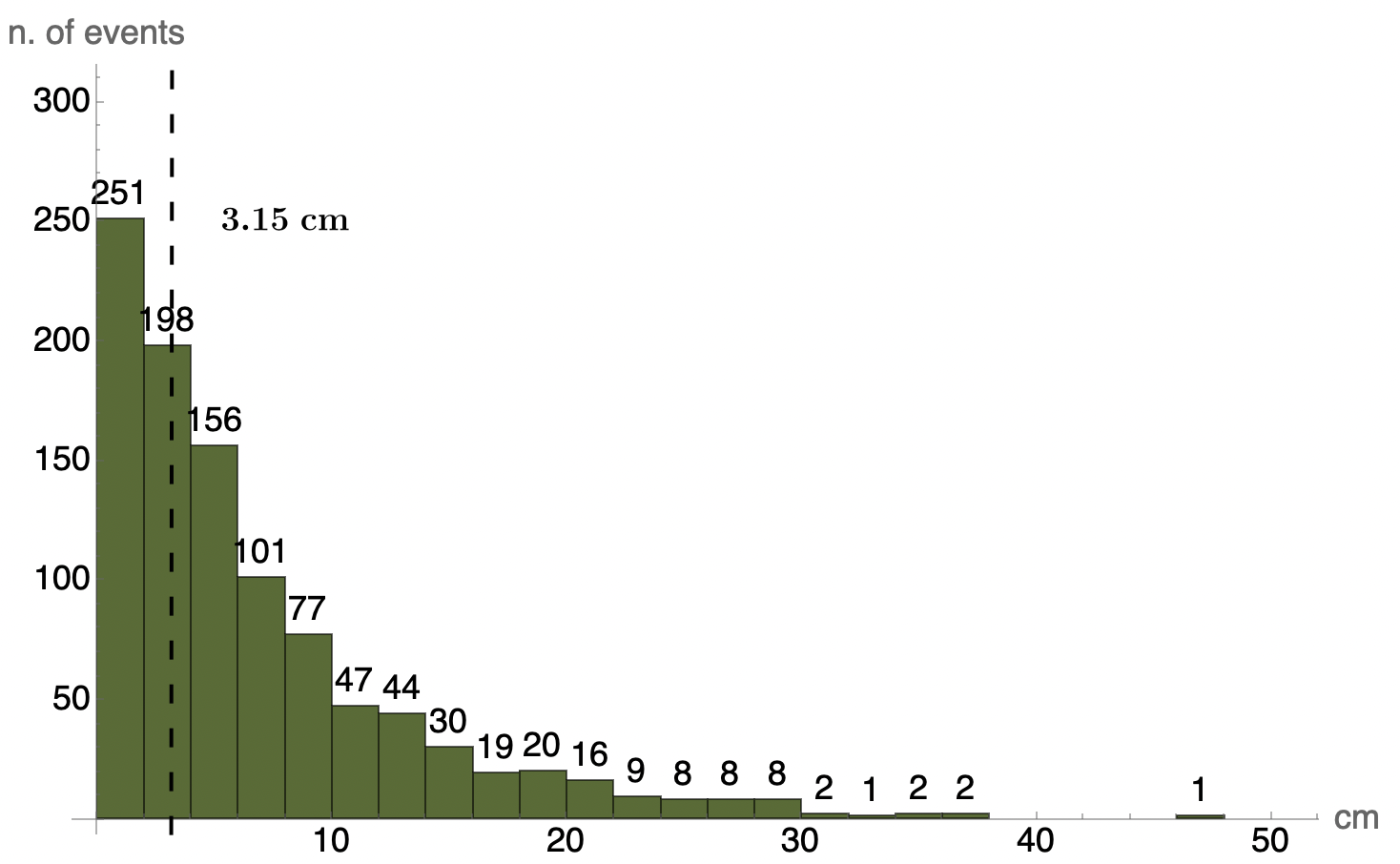}
\caption{\footnotesize Decay $\eta_c\to \Lambda \bar \Lambda$: Fraction (out of 1000) of $\Lambda$ baryons decaying at different lengths from the primary vertex. The vertical dashed line indicates where the inner surface of the beam pipe is located (3.15 cm away from the primary vertex).
\label{fig:penetration} 
}
\end{center}
\end{figure}

Taking into account the exponential dispersion of the decay times, about 58\% of the $\Lambda$ baryons coming out approximately at $90$ degrees decay inside the wall of the beam pipe or in the detector (see Fig.~\ref{fig:penetration}). Given a spherically symmetric cross section,  2/3  of all the scattered particles decay either  inside the wall of the beam pipe or in the MDC or the VELO. Even though the beam pipe walls are made of material that interacts the least possible with the particles, still they must be affected and, in particular, once inside the detector,  the spatial part of the density matrix is completely incoherent (because a spherical wave becomes a localized track in the detector). This picture is confirmed by the study of the cross section $\Lambda p \to \Lambda p$ in which the interaction between the hyperons and the beam-pipe wall is utilized~\cite{Dai:2022wpg,BESIII:2024geh}.

\begin{table}[h!]
\tablestyle[sansboldbw]
\begin{tabular}{*{3}{p{0.2\textwidth}}}
\theadstart
    \thead  &\thead mass (MeV) &
    \thead lifetime (s) \\
\tbody
 $\Lambda$ & 1115.683$\pm$ 0.006  & $(2.617 \pm 0.010)\times 10^{-10}$ \\
 $\Sigma^-$  &1197.449 $\pm$ 0.029    &  $(1.479 \pm 0.011) \times 10^{-10}$ \\
 $\Sigma^0$  &1192.642 $\pm$  0.024 &  $(7.4 \pm  0.7)\times 10^{-20}$ \\
 $\Xi^- $&  1321.71 $\pm$ 0.07& $(1.639 \pm 0.015) \times 10^{-10}$ \\
  $\Xi^0 $&  1314.86 $\pm$ 0.20& $(2.90 \pm 0.09) \times 10^{-10}$ \\
$\Omega^- $  &  1672.45 $\pm$ 0.29&  $(0.821\pm0.011) \times 10^{-10}$ \\
   \hline%
  \tend
\end{tabular}
\caption{\footnotesize \label{tab:baryons} \textrm{Masses and lifetimes of the strange baryons~\cite{ParticleDataGroup:2022pth}.}}
\end{table}

While only very weak interactions are expected between the spin of the $\Lambda$ baryons and the matter of the detector,  the density matrix as a whole should be affected by the loss coherence as the particle leaves a track in the detector~\cite{mott1929}.  
It does not seem so from the large entanglement shown by the data.    Why  are the spins of the $\Lambda$ baryons still entangled? 
 This problem does not appear to have been much discussed in the literature. As suggested in \cite{PhysRevA.89.032107}, whereas the spatial part becomes localized and describes the (feeble) track left by the $\Lambda$ baryon in the detector, the spin part remains (mostly) unchanged and still fully correlated. 
 
 The density matrix seems to factorize into  spatial- and  spin-dependent parts.
As already noticed in Section 4.1, this important feature can be put to the test in the case of  $\eta_c \to \Lambda \bar \Lambda$ and $\chi^0_c \to \Lambda \bar  \Lambda$. We know that the helicity amplitudes in vacuum are in this case fixed. If the measured ones turn out to be less than expected, it would  mean that some of the entanglement has been lost in the interaction with the detector. Vice versa, if the measured entanglement is maximal, the polarization density matrix must be factorized from the space dependent part.

The data from the decays of the $J/\psi$, $\psi(3686)$ and $\chi_{c}^{1}$ into different baryons cannot be readily utilized to test such a factorization. Even though the decay products do travel different distances outside the beam pipe, their form factors are different and the results cannot be compared.

A possible, albeit imperfect, test with the available data can be performed by taking the events from the direct production of $\Lambda$ baryons (as discussed in Appendix A) at different center-of-mass energies as given in~\cite{BESIII:2019nep} at $\sqrt{s}=2396$ MeV and in~\cite{BESIII:2023euh}  at $\sqrt{s}=$ 3680, 3683, 3684, 3685, 3687, 3691 and 3710 MeV. Fig.~\ref{fig:elist} shows the values of the coefficients $\alpha$ and $\Delta\Phi$ extracted from events at different values of $\sqrt{s}$. Figure~\ref{fig:lengths} shows the lengths travelled on average by the baryons with different energies.

The energy at threshold is sufficiently small to be well away from the $J/\psi$ very narrow resonance. It gives a value of the form factors uncontaminated by those of the $J/\psi$. Similarly, the last point at $\sqrt{s}=3710$ MeV is sufficiently away from the $\psi(3686)$ resonance---which is rather narrow too. We can take the values extracted for these two energies and compare them knowing that they come from baryons that have decayed inside the beam pipe (the first) and that have travelled through the wall of the beam pipe and into  the MDC detector (the second). Though the dependence of the form factors on the energy  may render the test unreliable, we can provisionally assume that the form factors have little variation with the energy when away from resonances.

Bearing in mind the above conditions, we can see that the values for the two limiting energies are close together, identical within one standard deviation. This we take as  a  circumstantial clue that the polarization density matrix is not affected by the loss of coherence that the spatial part is undergoing in the second case, where a faint trace of the baryon can be seen across the detector as a proof that superposition is no longer present. This is only a clue because the uncertainty in these values is large  and the value $\alpha=0$ cannot be excluded.

\begin{figure}[ht!]
\begin{center}
\includegraphics[width=3in]{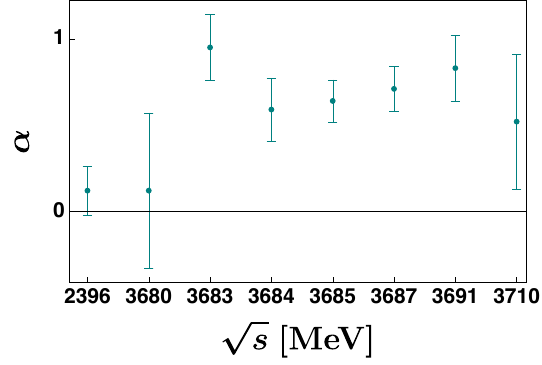}
\includegraphics[width=3.15in]{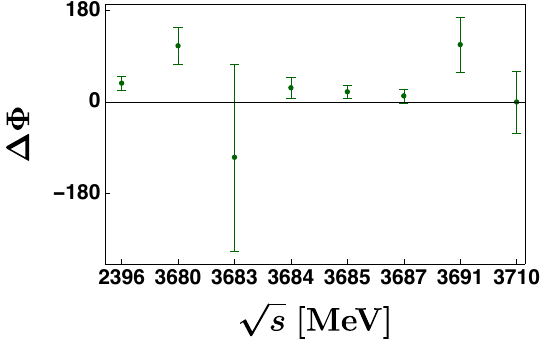}
\caption{\footnotesize Coefficients $\alpha$ and $\Delta\Phi$ for the production of $\Lambda \bar \Lambda$ at different values energies $\sqrt{s}$. The first point is just about threshold, the others around the $\psi(3686)$ mass. 
\label{fig:elist} 
}
\end{center}
\end{figure}

\begin{figure}[ht!]
\begin{center}
\includegraphics[width=3in]{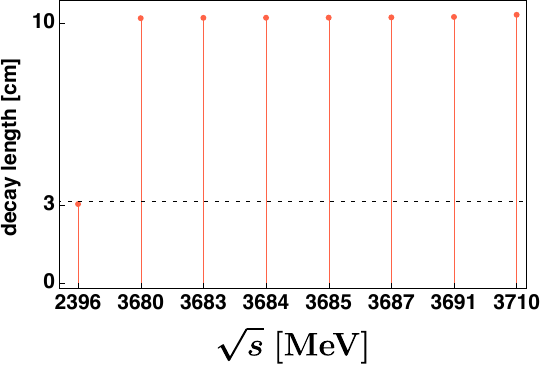}
\caption{\footnotesize Decay lengths for the $\Lambda$ baryons for the different values of $\sqrt{s}$. The first track is  just inside the beam pipe (with an inner wall at 3.15 cm, marked by the dashed horizontal line), the last goes through the beam pipe and into the MDC (which stands between a radius of 5.9  and  81 cm).
\label{fig:lengths} 
}
\end{center}
\end{figure}

A clear cut discussion for the data from the decays of $J/\psi$ and $\psi(3686)$ is not possible because of  their  different form factors---on which the final entanglement depends. For these decays, it would be useful to have the helicity amplitudes computed on separate sets of baryons, namely those for which the decay takes place inside the beam pipe and those for which it takes place in the detector.

\section{Loopholes at colliders}

{\versal The experimental violation} of the Bell inequality at low energies has been challenged by invoking the presence of loopholes that bypass  its effect. Though the same loopholes might be brought to bear to tests at high energies, their effectiveness and significance is different in the new settings and must be revisited.

For a start, two of them, namely the detection and coincidence loopholes, do not seem to apply  at colliders.  The first because one routinely assumes  having a fair sampling of the recorded events. This assumption is necessary given the small fraction of actually recorded events out of the many produced.  
Also, the coincidence loophole does not seem to be problematic at colliders. Possible misidentifications are always accounted for in the quoted uncertainty in the results of the experiments. 

 Next, the locality loophole seems to be possible. It is potentially present for states made of particles  that end up decaying with a relative time-like interval, either because they decayed at different times or because they do not move apart fast enough. To close the locality loophole it is desirable to consider decays in which the produced particles are identical,  and therefore their lifetimes are also the same. Even in this case, the actual decays take place with an exponential spread. To take this into account, one must verify that the majority of the events do take place  separated by a space-like interval and weed out those that do not.
 The selection of these events could be implemented with a suitable cut on the relative momentum of the two particles. If the amount of available data is large and the  fraction of pairs rejected  by the cut is small, this refinement would not affect the significance of the Bell test under consideration.

Finally,  the polarization measurement is made inside the detector by the particles themselves as they decay into the final state; because the decay is a quantum process, it is the ultimate random generator and  therefore the freedom-of-choice  loophole is addressed. 

In considering the relevance of these loopholes, the reader should bear in mind that there is no model based on hypothetical deterministic variables that is able to explain all the experimental tests by exploiting one or more of these loopholes.  We can now add to the list of these tests those performed at high energies. At low energy all the possible loopholes have been closed~\cite{Hensen:2015ccp,Giustina:2015yza,Storz:2023jjx}.

\vskip1cm
\section{Outlook}

{\versal We have established} that quantum entanglement and  the violation of Bell inequality are both present in several decays of charmonium states. The charmonium offers the ideal laboratory for the study of these properties exclusive to quantum systems. If the experimental Collaborations were to provide some of the missing information on the helicity amplitudes  of the processes we have discussed, namely phases  and correlations  in the uncertainty, tests even tighter than those hereby discussed  could be established.

 \appendix
 
 \section{$\Lambda\text{-}\bar \Lambda\; \text{baryon pair production}$}

 {\versal A pair of  $\Lambda$ baryon} and anti-baryon can be produced directly  if the energy in the center of mass is below the mass of the first charmonium bound state, that is roughly below 3 GeV.
In this process the $\Lambda$ baryons are produced directly through their coupling to the photon 
\be
e^{+} + e^{-} \to \Lambda + \bar \Lambda\,,
\ee
without going through an intermediate resonance. The helicity amplitudes depends on the electromagnetic form factors of   the baryon, which enters the current, as in \eq{current}, as
\be
\bar u_{\Lambda} \left[ F_{1} \gamma^{\mu} + \frac{1}{2 m_{\Lambda}} \sigma^{\mu\nu} q^{\nu} F_{2}\right] u_{\Lambda} A_{\mu}\label{current2}
\ee
in which  $A_{\mu}$ stands for the electromagnetic four vector and $q^{2}= s$ is the center-of-mass energy. The other common parameterization, the one we used in Section 4.6, is
in terms of the form factors
\be
G_{M} = F_{1}+F_{2} \quad \text{and} \quad G_{E} = F_{1} + \frac{q^{2}}{2 m_{\Lambda}^{2}}F_{2}\,,
\ee
which is obtained from \eq{current} by means of the Gordon decomposition.

\begin{figure}[ht!]
\begin{center}
\includegraphics[width=3in]{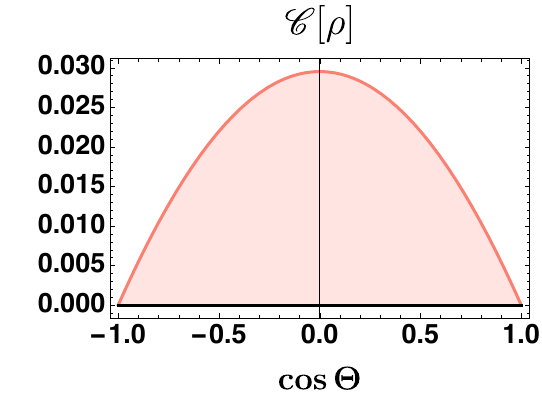}
\includegraphics[width=3.15in]{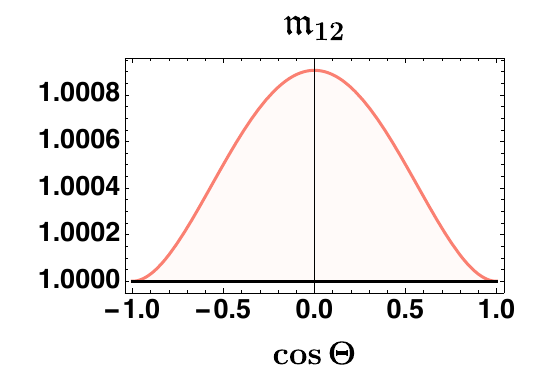}
\caption{\footnotesize Concurrence (left) and Horodecki's condition $\mathfrak{m}_{12}$ (right) for $e^+ e^- \to \Lambda \xbar{\Lambda}$.
\label{fig:ee} 
}
\end{center}
\end{figure}

To make contact with the experiments, we rewrite the two form factors  $G_{M}$ and $G_{E}$ in terms of the two coefficients $\alpha$ and $\Delta \Phi$, introduced in Section 4.

The analysis of the data  taken at $\sqrt{s} = 2.396$ GeV  gives~\cite{BESIII:2019nep}
\be
\alpha = 0.12\pm0.14|_{\rm stat} \quad \text{and} \quad \Delta \Phi =  0.65\pm0.21|_{\rm stat}\, ,
\ee
in which the two statistical uncertainties are correlated with coefficient 0.17.

As discussed in Section 6, the data collected are for events with energies  just above the threshold  for the production of the $\Lambda$ baryons and therefore the baryons move slowly, decaying within about 2.4 cm from the production point. These decays take place within the beam pipe and before any  interaction with the detector is possible.

The analysis of entanglement and Bell inequality violation is completely analogous to that in Sec.~\ref{sec:1to1212}; at $\Theta=\pi/2$, we find 
\be
\mathscr{C} [\rho] =0.12 \pm0.11 \quad \text{and} \quad \mathfrak{m}_{12} =1.01 \pm 0.04\, .
\ee
 The Bell inequality is  not violated.

The analysis of the data  taken at $\sqrt{s} = 3.710$ GeV  gives~\cite{BESIII:2023euh}
\be
\alpha = 0.52^{+0.38}_{-0.39}|_{\rm stat}\pm0.02|_{\rm syst} \quad \text{and} \quad \Delta \Phi =  0.0^{+1.13}_{-0.99}|_{\rm stat}\pm0.03|_{\rm syst}\, ,
\ee

Now the data collected are for events in which the decay takes place about 10 cm  from the production point, that is, after the $\Lambda$ baryons have crossed most of the MDC detector, let alone the beam pipe. 

At $\Theta=\pi/2$, we find
\be
\mathscr{C}[\rho] = 0.52\pm0.30 \quad \text{and} \quad \mathfrak{m}_{12} = 1.27\pm0.61 \, .
\ee
 The Bell inequality is  not violated.

A similar analysis can be done on $e^+e^-\to\Sigma^+\bar \Sigma^-$ data~\cite{BESIII:2020uqk}.

 \section[$\Lambda_b \to J/\psi +  \Lambda$]{$\boxed{\Lambda_b \to J/\psi +  \Lambda}$}
 
{\versal The helicity states} of the final system in
\be
\Lambda_b \to J/\psi +  \Lambda
\ee
fall in the $\tfrac{1}{2}$ representation of the product $1 \otimes \tfrac{1}{2} = \tfrac{3}{2} \oplus \tfrac{1}{2}$. It is constrained by the conservation of the angular momentum to be described by the two states
\bea
|\psi_\uparrow\rangle &\propto & w_{0\,- \frac{1}{2}} \,| 1 ,0 \rangle \otimes |\tfrac{1}{2}, \tfrac{1}{2} \rangle +
w_{1\,\frac{1}{2} }\, |1, 1 \rangle  \otimes |\tfrac{1}{2}, -\tfrac{1}{2} \rangle \label{states1}\\
|\psi_\downarrow\rangle &\propto & w_{0\, \frac{1}{2}} \,| 1 ,0 \rangle \otimes |\tfrac{1}{2}, -\tfrac{1}{2} \rangle +
w_{-1\,-\frac{1}{2} }\, |1, -1 \rangle  \otimes |\tfrac{1}{2}, \tfrac{1}{2} \rangle\, , \label{states2}
\eea
in which the state in \eq{states1} corresponds to the $\Lambda_b$ with positive helicity and that in \eq{states2} to the opposite helicity.

The two states in \eqs{states1}{states2}  enter, depending on the polarization $P_b$ of the initial $\Lambda_b$, in the mixture
\be
\rho_{\Lambda\,J/\psi} \propto p_\uparrow \, |\psi_\uparrow\rangle \langle \psi_\uparrow| +
p_\downarrow  \, |\psi_\downarrow\rangle \langle \psi_\downarrow| \, ,
\ee
in which $p_\uparrow= \tfrac{1}{2} + P_b$ and $p_\downarrow=  \tfrac{1}{2} -P_b$. 

There is no parity conservation and therefore 4 independent non-vanishing helicity amplitudes: $w_{_{0\, \pm \frac{1}{2}}}$ and $w_{_{\pm 1\, \pm \frac{1}{2}}}$. The  density matrix is
\be
\small
\rho_{\Lambda\,J/\psi} \propto \begin{pmatrix} 
 |w_{-1\, -\frac{1}{2}}|^2 & 0 &0 & w_{-1\,-\frac{1}{2} }w^*_{0\, \frac{1}{2}} &0 &0 \\
0 & 0& 0&  0&0 &0\\
0 & 0& |w_{0\, -\frac{1}{2}}|^2 &  0& 0 & w_{0\,\frac{1}{2} }w^*_{1\, \frac{1}{2}} \\
w_{0\,  \frac{1}{2}} w^{*}_{-1\, -\frac{1}{2}}  &  0 & 0 &|w_{0\, -\frac{1}{2}}|^2 & 0&0 \\
0&   0& 0&0&0 & 0 \\
0 & 0& w_{1\,  \frac{1}{2}} w^{*}_{0\, -\frac{1}{2}} & 0  &0 &  |w_{1\, \frac{1}{2}}|^2 
\end{pmatrix}\, ,
\ee
with no angular dependence.
The helicity amplitude $w_{_{1\, \frac{1}{2}}}$ is expected to be small because of the mostly chiral coupling.

Data from one of the experimental collaborations~\cite{ATLAS:2014swk} have been taken for the chain decay $\Lambda_{b}\to J/\psi (\to \mu^{+}\mu^{-}) \, \Lambda (p\pi^{0})$ with a luminosity of 4.6 fb$^{-1}$ at center-of-mass energy of 7 TeV and recorded by the ATLAS detector at the LHC. They extract from a likelihood fit the helicity amplitudes:
\bea
|w_{_{0\,  \frac{1}{2}}} |=  0.17^{+0.12}_{-0.17}|_{\rm stat}\pm 0.09|_{\rm syst}\, , &\quad &|w_{_{0\,  -\frac{1}{2}}} |= 0.59^{+0.06}_{-0.07}|_{\rm stat}\pm 0.03|_{\rm syst}\nn \\
|w_{_{1\,  \frac{1}{2}}}|=  0.08^{+0.13}_{-0.08}|_{\rm stat}\pm 0.06|_{\rm syst}&\quad & |w_{_{-1\,  -\frac{1}{2}}}|= 0.79^{+0.04}_{-0.05}|_{\rm stat}\pm 0.02|_{\rm syst}\, .
\eea
Data for the same amplitudes are available from CMS~\cite{CMS:2018wjk} and LHC$b$~\cite{LHCb:2013hzx}.

We find a positive but small value:
\be
{\cal N} (\rho) = 0.05\pm0.06 \, ,
\ee
which corresponds to essentially no entanglement within one standard deviation.  This means that the final state is a mixture of the two states in
\eqs{states1}{states2} with equal weight---which is what we would expect if the $\Lambda^b$ is produced unpolarized.

\section*{Acknowledgements}
{\small
We thank Isabella Garzia for help in understanding the geometry of the BESIII beam pipe and detector.  L.M. is supported by the Estonian Research Council grants PRG803, RVTT3 and by the CoE program grant TK202 ``Fundamental Universe'’.}

\begin{multicols}{2}
\small
\bibliographystyle{JHEP}   
\bibliography{charmonium.bib} 

\end{multicols}
\end{document}